\documentclass{emulateapj}

\def \h2         {\hbox{H$_2$}}

\def\approxlt{\lower.2em\hbox{$\buildrel < \over \sim$}}
\def\approxgt{\lower.2em\hbox{$\buildrel > \over \sim$}}

\def \aap        {{ApJ}}

\begin{document}

\title {Eddington-Limited Accretion in \MakeLowercase{z}$\sim2$ WISE-selected Hot, Dust-Obscured Galaxies}

\author {Jingwen Wu\altaffilmark{1,2},  
Hyunsung D. Jun\altaffilmark{3,4},
Roberto J. Assef\altaffilmark{5},
Chao-Wei Tsai\altaffilmark{2}, 
Edward L. Wright\altaffilmark{2},
Peter R. M. Eisenhardt\altaffilmark{3},
Andrew Blain\altaffilmark{6},
Daniel Stern\altaffilmark{3},
Tanio D\'iaz-Santos\altaffilmark{5},
Kelly D. Denney\altaffilmark{7,8},
Brian T. Hayden\altaffilmark{9,10},
Saul Perlmutter\altaffilmark{9,10},
Greg Aldering\altaffilmark{9},
Kyle Boone\altaffilmark{9,10},
Parker Fagrelius\altaffilmark{9,10}
}

\altaffiltext{1}{National Astronomical Observatories, Chinese Academy of Sciences, 20A Datun Road, Chaoyang District, Beijing, 100012, China; jingwen@nao.cas.cn}
\altaffiltext{2}{Department of Physics and Astronomy, University of California, Los Angeles, CA 90095, USA}
\altaffiltext{3}{Jet Propulsion Laboratory, California Institute of Technology, 4800 Oak Grove Dr., Pasadena, CA 91109, USA}
\altaffiltext{4}{School of Physics, Korea Institute for Advanced Study, 85 Hoegiro, Dongdaemun-gu, Seoul 02455, Korea}
\altaffiltext{5}{Universidad Diego Portales, Av Republica 180, Santiago, Regi\'on Metropolitana, Chile}
\altaffiltext{6}{Department of Physics and Astronomy, University of Leicester, LE1 7RH Leicester, UK}
\altaffiltext{7}{Department of Astronomy, The Ohio State University, 140 West 18th Avenue, Columbus, OH 43210, USA}
\altaffiltext{8}{Illumination Works, LLC, 5550 Blazer Pkwy, Dublin, OH, 43017, USA}
\altaffiltext{9}{E.O. Lawrence Berkeley National Lab, 1 Cyclotron Road., Berkeley, CA, 94720, USA}
\altaffiltext{10}{Department of Physics, University of California Berkeley, Berkeley, CA 94720, USA}

\begin{abstract}

Hot, Dust-Obscured Galaxies, or ``Hot DOGs", are a rare, dusty, hyperluminous galaxy population discovered by the WISE mission. Predominantly at redshifts 2-3, they include the most luminous known galaxies in the universe. Their high luminosities likely come from accretion onto highly obscured super massive black holes (SMBHs). We have conducted a pilot survey to measure the SMBH masses of five $z\sim2$ Hot DOGs via broad H$\alpha$ emission lines, using Keck/MOSFIRE and Gemini/FLAMINGOS-2. We detect broad H$\alpha$ emission in all five Hot DOGs. We find substantial corresponding SMBH masses for these Hot DOGs ($\sim 10^{9} $M$_{\odot}$), and their derived Eddington ratios are close to unity.  These $z\sim2$ Hot DOGs are the most luminous AGNs for their BH masses, suggesting they are accreting at the maximum rates for their BHs. A similar property is found for known $z\sim6$ quasars. Our results are consistent with scenarios in which Hot DOGs represent a transitional, high-accretion phase between obscured and unobscured quasars. Hot DOGs may mark a special evolutionary stage before the red quasar and optical quasar phases, and they may be present at other cosmic epochs.

\end{abstract}


\section{Introduction} \label{intro}

Gravitational accretion onto super massive black holes (SMBHs) is one of the major energy production sources in galaxies, powering active galactic nuclei (AGN). Its scale and intensity (generally described as the quasar luminosity normalized by the Eddington luminosity, linearly proportional to the mass of the black hole, i.e. the ``Eddington ratio") is thought to be the most important parameter to govern AGN properties (e.g., Trump et al. 2011, Shen \& Ho 2014). High Eddington ratios (as well as super Eddington accretion) have only been reported in a small fraction of unobscured quasars (e.g., Shen et al. 2008, 2011; Jun \& Im 2013).  The average Eddington ratio for all SDSS quasars is 0.26 $\pm$ 0.34 (Shen et al.  2011). 
Systematically higher Eddington ratios, slightly above unity, have been reported for SMBHs in quasars at very high redshifts ($z \sim 6$ and above; Fan et al. 2006, Willott et al. 2010, De Rosa et al. 2011), while in the local universe, ultraluminous X-ray sources (ULXs) are believed to often be hosted by stellar mass compact objects accreting significantly above the Eddington limit.  Notably, X-ray pulsations have been detected for three ULXs, implying the compact object in those systems are neutron stars (Bachetti et al. 2014, F\"urst et al. 2016, Israel et al. 2017a), with an inferred peak isotropic luminosity of $\sim 1000$ times the Eddington limit in at
least one system (Israel et al. 2017b).
Despite its importance, the observed Eddington limited accretion in $z=6-7$ quasars  remains poorly understood. It may be the key to understanding the formation and co-evolution of the first generation of SMBHs and their host galaxies, and to explaining how a few to 10 billion solar mass SMBHs were able to build up their masses when the universe is only $\sim$ 1 Gyr old (e.g., Wu et al. 2015).

Explanations for the coevolution of SMBHs and their hosts often focus on galaxy merging  (e.g., Hopkins et al. 2006, 2008, Somerville et al. 2008). In these scenarios, 
major mergers trigger a starburst at their coalescing center, feed SMBHs, whose growth and feedback will eventually sweep out gas and dust, terminating star-formation and leaving a visible quasar with fading luminosity. During these processes, highly obscured, highly luminous phases accompanied by highly accreting SMBHs are expected. These models predict an important phase where galaxies areÊ highly obscured, yet exceedingly luminous.

High luminosity, infrared dominant galaxies were first discovered by Kleinmann \& Low (1970) and Rieke \& Low (1972). SignificantÊ populations of such galaxies were found by the {\it Infrared Astronomical Satellite} ({\it IRAS}, Neugebauer et al. 1984) in the local universe, then extended to the more distant universe (especially to the most rapid evolution epoch at $z=2-3$) by later infrared space missions and ground-based submillimeter facilities. Milestones include the discovery of the submillimeter galaxies (SMGs, see Blain et al. 2002 and Casey et al. 2014 for a review) in 850 $\mu$m and 1 mm surveys, and the discovery of dust-obscured galaxies (DOGs, Dey et al. 2008) by {\it Spitzer} 24$\mu$m surveys (e.g., Rigby et al. 2004, Donley et al. 2007, Yan et al. 2007, Farrah et al. 2008, Lonsdale et al. 2009). Both populations are at $z \sim2$ (Chapman et al. 2005, Dey et al. 2008).
 In general, SMGs are thought to be at an earlier phase of the merger/AGN evolution when the starburst still dominates the luminosity, while in the DOG phase, a hidden AGN becomes increasingly influential, representing a transitional stage when the AGN contribution overtakes star formation and becomes the dominant energy source (Dey et al. 2008, Bussmann et al. 2009, Narayanan et al. 2010). 
More recently, surveys with the {\it Herschel Space Telescope} (Pilbratt et al. 2010), as well as the South Pole Telescope (Carlstrom et al. 2011) and the Atacama Cosmology Telescope (Swetz et al. 2011) have extended the area surveyed at submillimeter wavelengths to hundreds of square degrees, highlighting dusty star forming galaxies  to even higher redshifts (e.g., Riechers et al. 2013, Dowell et al. 2014), although a large fraction of the most luminous galaxies are lensed (e.g., Negrello et al. 2010, Vieira et al. 2013, Bussmann et al. 2015). The WISE mission (Wright et al. 2010) enabled all-sky selection of dusty, infrared galaxies, and is more efficient in selecting red quasars than previous surveys (e.g. Glikman et al. 2007), either by WISE colors alone (e.g., Stern et al. 2012, Assef et al. 2013), or combined with other UV to near-IR surveys (e.g., Banerji et al. 2013, 2015, Hainline et al. 2014). 

Benefiting from its all-sky coverage, WISE is able to identify the most extreme, dusty, highly obscured AGNs in the universe. One of the most successful examples is the discovery of the hyperluminous, hot, dust-obscured galaxies (Hot DOGs, Eisenhardt et al. 2012, Wu et al. 2012). These galaxies were selected by looking for objects strongly detected by WISE at 12 and/or 22 $\mu$m, but only faintly or not at all at 3.4 and 4.6 $\mu$m, i.e.  ``W1W2drop-out" galaxies (Eisenhardt et al. 2012). The selected galaxies lie between redshifts 1.5 and 4.6, their distribution peaking at $z=2-3$ (Eisenhardt et al. 2012, Assef et al. 2015). Their luminosities are high (Wu et al. 2012, Bridge et al. 2013, Jones et al. 2014, Assef et al. 2015, Tsai et al. 2015, Fan et al. 2016a, Farrah et al. 2017), mostly well above 10$^{13}$ L$_{\odot}$, and they do not show evidence of lensing (Wu et al. 2014, Tsai et al. 2015). The most luminous 10\% even exceed 10$^{14}$ L$_{\odot}$, comparable to the most luminous quasars known (Assef et al. 2015), and include the single most luminous galaxy or quasar on record so far, the Hot DOG WISE J2246-0526 at $z=4.593$ (Tsai et al. 2015, Diaz-Santos et al. 2016). This discovery achieved one of the primary WISE science goals: finding the most luminous galaxies in the universe. 

Having a consistent spectral energy distribution (SED) shape (Wu et al. 2012, Tsai et al. 2015, Tsai et al.  in prep), which is characterized by an unusually high mid-IR to submm ratio, Hot DOGs are quite different from most other well-studied IR luminous populations: they contain greater proportions of hot dust, with a characteristic dust temperature of 60-100 K (Wu et al. 2012, Bridge et al. 2013, Fan et al. 2016a), significantly hotter than the typical 30-40 K observed in SMGs or DOGs (e.g., Melbourne et al. 2012, Magnelli et al. 2012). They are much brighter and rarer than DOGs (there are about 0.03 deg$^{-2}$ vs. 300 deg$^{-2}$). Although not required,  for selection using WISE colors, they all satisfy the DOG selection criteria (Dey et al. 2008) but are much hotter; therefore we have dubbed them as 'Hot DOGs' (Wu et al. 2012). 

Hot DOGs likely host very powerful AGNs, as indicated by their high dust temperatures and SEDs, and the AGN  dominates their luminosities (Eisenhardt et al. 2012, Assef et al. 2015, Tsai et al. 2015, Fan et al. 2016a, Farrah et al. 2017). These AGNs have very high extinction, typically $A_{V}$ $\sim$ 20 and up to $A_{V} \sim$ 60 (Assef et al. 2015),  and are close to Compton-thick in the X-ray bands (Stern et al. 2014, Piconcelli et al. 2015, Assef et al. 2016, Vito et al. in prep.).  They are likely experiencing very strong feedback: a significant number of such galaxies present extended Lyman-$\alpha$ blob (LABs) structures extended over 10's of kpc (Bridge et al. 2013). A recent high spatial resolution [CII] observation with ALMA of the most luminous Hot DOG W2246-0526 reveals an extended, uniform, highly turbulent ISM, indicative of an isotropically expelling  galaxy-scale event (D\'iaz-Santos et al. 2016).  

Their high luminosities, hot dust temperatures, and strong feedback suggest Hot DOGs may be in transition between the obscured and unobscured phases of luminous quasars, when the surrounding dust and gas are being heated and blown out, just before visible quasars emerge (Wu et al. 2012, Bridge et al. 2013, Assef et al. 2015, D\'iaz-Santos et al. 2016, Fan et al. 2016b). Given the high luminosities for Hot DOGs, the inferred BH mass must be well above the local BH mass-host galaxy correlation if a typical AGN accretion rate is assumed, or, alternatively, the Eddington ratio must be very high, even above the Eddington limit (Assef et al. 2015, Tsai et al. 2015). Potentially both factors may be present. Therefore, a measurement of their BH masses is a key step to understand Hot DOGs. 

Here we present a pilot survey to measure the BH mass of five Hot DOGs at $z \sim2$ using the line width of the broad H$\alpha$ line redshifted into the near-IR. A description of the observations and data reduction is given in Section 2, and in Section 3 we fit the line width of the spectra. The estimated BH masses, luminosities, and Eddington ratios are presented in Section 4. In Section 5, we compare Hot DOGs to other galaxy populations at $z \sim2$, as well as to quasars at $z \sim6$. Conclusions are summarized in Section 6. Throughout this paper, we assume a $\Lambda$CDM cosmology with  $H_0 = 70$ km s$^{-1}$ Mpc$^{-1}$, $\Omega_m = 0.3$, and $\Omega_{\Lambda} = 0.7$ .

\section{Observations and Data Reduction} \label{obs}

We observed five Hot DOGs with secure redshifts and well sampled SEDs. All selected targets have redshifts  $z\sim 2$ so their H$\alpha$ line is observable with near-IR spectroscopy. Except for requiring coordinates that made them accessible during the observing runs, Êno other selection criteria were applied when choosing these targets. Their extinctions have been calculated from a model of their rest-frame UV-to-mid-IR SEDs that includes a starburst, evolved stars, and reddened AGN components (Assef et al. 2015). The derived extinctions  range from $A_{V}\sim 8$  to 25.  Source information is listed in Table 1.

Normally at such high extinction, broad line regions (BLRs) will be hard to see. But for some Hot DOGs, we see hints of BLR emission even in the rest-UV (Wu et al. 2012, Assef et al. 2016).  One possible explanation is that  AGNs in these Hot DOGs are so luminous compared to their host galaxies, BLR emission leaks through the torus, or perhaps due to reflection of BLR emission by dust.  A detailed discussion of this leaking/scattering explanation for Hot DOGs is given in Assef et al. (2016), in which a subsample of Hot DOGs has been found to have excess UV/optical emission (blue excess Hot DOGs, or ``BHDs"), which can be best explained as unobscured light leaked from the AGN by reflection off dust.  

\subsection{Keck/MOSFIRE}

We obtained near-IR spectroscopy for four Hot DOGs (WISE J033851.33+194128.6, WISE J090439.84+394715.2, WISE J113634.29$+$423602.9 and WISE J213655.74$-$163137.8 (hereafter W0338+1941, W0904+3947, W1136+4236 and W2136$-$1631, respectively) using the Multi-Object Spectrometer For Infrared Exploration (MOSFIRE) instrument (McLean et al. 2010, 2012) at Keck Observatory. $K$-band spectra for three Hot DOGs and $H$-band spectra for two Hot DOGs were obtained during three runs in 2014 and 2015 (one target was observed in both bands). The MOSFIRE $K$-band filter is centered at 2.162 $\mu$m with a full-width at half-maximum (FWHM) of 0.483 $\mu$m. The $H$-band filter is centered at 1.637 $\mu$m with a FWHM of 0.341$\mu$m. The data
were acquired using masks with a slit width of 0.7$\arcsec$, giving a velocity resolution of  $\Delta v \sim$ 80 km s$^{-1}$ ($R \sim 3600$). Each individual exposure time was 2$-$3 mins. The observation dates and total integration times are listed in Table 1.  The seeing during the observations was $\sim$0.5$\arcsec$ for W2136-1631, and 1$\arcsec$-1.2$\arcsec$ for the rest of the targets. 

The data were reduced using the MOSFIRE data reduction pipeline (DRP), which performs flat fielding, sky subtraction and wavelength calibration, and outputs rectified two-dimensional (2D) spectra, from which one-dimensional spectra were extracted. Telluric correction and flux calibration were performed using the spectra of A0 standard stars.

\subsection{Gemini/FLAMINGOS-2}
WISE J221648.05+072353.6 (hereafter W2216+0723) was observed using  FLAMINGOS-2 (Eikenberry et al. 2012) at the Gemini-South Observatory on the night of UT 2014 November 7. It was observed
simultaneously in the $J$ and $H$ bands using the $JH$ grating with the 2
pixel-wide (0.36\arcsec) longslit, providing a spectral resolving
power of approximately $R \sim 1000$. The target was observed for 1 hr in an
ABBA sequence. Additional observations were obtained on the nights of
UT 2014 June 2, UT 2014 July 13 and UT 2014 November 6 using the same
configuration but under much poorer weather conditions, so we do not
consider them any further in this analysis.

The data were reduced using standard
{\tt{IRAF}}\footnote{\url{http://iraf.noao.edu}} tools with the
sky emission lines used for wavelength calibration. The star HIP106817 was
used for telluric and flux calibration using the
${\tt{XTellCorr\_General}}$ routine of the {\tt{Spextool}} package
(Vacca, Cushing \& Rayner 2003).

\section{Line width fitting} \label{results}

We detected broad H$\alpha$ lines in all five Hot DOGs. We modeled the lines using the IDL routine MPFIT (Markwardt 2009).
The aim of the fitting is to measure the line width of the H$\alpha$ line originating from the BLR which is broadened by virial motion, in order to estimate the black hole mass (see section 4.1). There are two major concerns that can affect the fitting result significantly: one is the blending with emission lines from the narrow line regions (NLRs), the other is any contribution from outflows. We discuss them separately next. 

\subsection{ Fitting  spectra with broad lines and narrow lines} 
The broad H$\alpha$ line is blended with narrow H$\alpha$ and [NII]$\lambda$6549, 6585 emission, which need to be removed when their contributions are not negligible. We fit the broad line H$\alpha$, and narrow lines H$\alpha$, [NII]$\lambda$6549, 6585,  [OI]$\lambda$6300, 6364 and [SII]$\lambda$6717, 6732 simultaneously. We fix the line ratios for the [NII]$\lambda$6549, 6585 doublets to be 1:2.96 according to Greene \& Ho (2005), in order to decompose the  possible blending between the H${\alpha}$ and [NII] doublets atÊ relatively low SNR. We fit the rest-frame 6000-7000~\AA\ spectra including a power-law continuum and Gaussian lines, where all the components are limited to have non-negative fluxes and reported line widths are corrected for instrumental resolution.  The modeled spectra are plotted in Figure 1.  We iterated the fit once to include just the inner 98.76\% of the data sorted in absolute value of the residual (2.5$\sigma$ Gaussian rejection), and minimized $\chi^{2}_\nu$ over the 6000-7000~\AA\ range.

In this Section, we focus on the question of whether the broad line and/or narrow line components are necessary to fit the near-IR spectra. We explored three strategies to select the best parametrization to fit the spectra. In strategy 1, we only fit one broad Gaussian (FWHM $> $1000 km s$^{-1}$) for H$\alpha$, assuming the narrow H$\alpha$ and [NII] are negligible compared to the broad H$\alpha$ (``1B alone" strategy hereafter). 

In strategy 2, we fit a single narrow Gaussian (FWHM $<$1000 km s$^{-1}$) in addition to a single broad Gaussian to H$\alpha$ (``1B+1N" hereafter). We kept all narrow line widths the same and allowed the broad component's center to range within $\pm$1000 km s$^{-1}$ from the narrow H$\alpha$ redshift (e.g.,  Bonning et al.  2007,  Shen et al. 2011), determined from the peak of the narrow line model fit. 

In strategy 3, we allow two broad Gaussians and one narrow Gaussian to fit H$\alpha$ (``2B+1N" hereafter), with the reported FWHM of H$\alpha$ derived by the combination of the two broad Gaussians. The second broad component is assumed to have the same redshift as the narrow lines. Combining multiple Gaussians to obtain the broad FWHM has been used in many other works (e.g., Greene \& Ho 2005, Assef et al. 2011, Jun et al. 2017).  We present the fitted spectra in Figure 1. The fitting parameters including FWHM, the errors from MPFIT, $\chi^{2}_{\nu'}$ ($\chi^{2}_{\nu}$ calculated only covering H$\alpha$ wavelengths: 6450-6650~\AA),  and the degrees of freedom (DOF) of the best fit model for each strategy are listed in Table 2.  

To determine which strategy works best, we applied an F-test  to each spectrum over rest-frame 6450-6650~\AA. The results are listed in Table 3. The values in Table 3 show the probability  $p$ that the $\chi^{2}_\nu$ of the fit with the larger number of components was consistent with being drawn from the same distribution as the fit with the smaller number of components. Generally, when $p< 0.05$, we can assume that an extra component is necessary in the fitting. Note that when $p=0$, it means that the number is so small that it is below the numerical precision of the integrator we used.

Based on Table 2 and Table 3, the option 1 ``1B alone" strategy is ruled out by the F-test analysis, for all five targets, implying the narrow line component is not negligible. [A ``1 narrow line alone" (1N alone) case vs. ``1B + 1N"  is also rejected by the F-test]. Both broad line and narrow line components are necessary to represent the spectra.  

As seen in Figure 1,  W0338+1941 and W0904+3947 have poorer signal to noise ratios (SNRs) than the other three targets. According to the F-test,  the 2B+1N model is strongly preferred in two of the three high SNR targets. Since all Hot DOGs are arguably physically similar,  we assumed that a consistent model should apply to all targets. In this paper we use the 2B+1N model to fit all sources. For the one high SNR Hot DOG (W2216+0723) where the 2B+1N model is not preferred over the 1B+1N model by the F-Test,  the resulting FWHMs and implied BH masses for the two models are consistent (see Table 2).Ê

We adopt the FWHM and $\chi^{2}_\nu$ from the 2B+1N model in MPFIT as best-fit values for each target, then we estimate the error of this best-fit FWHM using a Monte Carlo approach similar to that of Assef et al. (2011). For a given spectrum, we first scale the uncertainty of each pixel such that the $\chi^{2}_\nu$ of the best fit model is equal to 1. We then create 1,000 simulated spectra with the same pixel size as the observed 
spectrum. The value of each pixel in the simulated spectra is randomly drawn from a Gaussian distribution centered at the flux of the best-fit model in that pixel, with a dispersion equal to its scaled uncertainty. We then fit each of the simulated spectra in the same way as described above. The 68.3\% confidence interval of the FWHM is obtained from the distribution of the best-fit FWHM to the simulated spectra.  Specifically, the range contains 68.3\% of the FWHM values below and above the median of the distribution. This uncertainty is listed in Table 4 as the asymmetric 1 $\sigma$ error of the FWHM.

\begin{figure*}
\epsscale{0.9}
\plotone{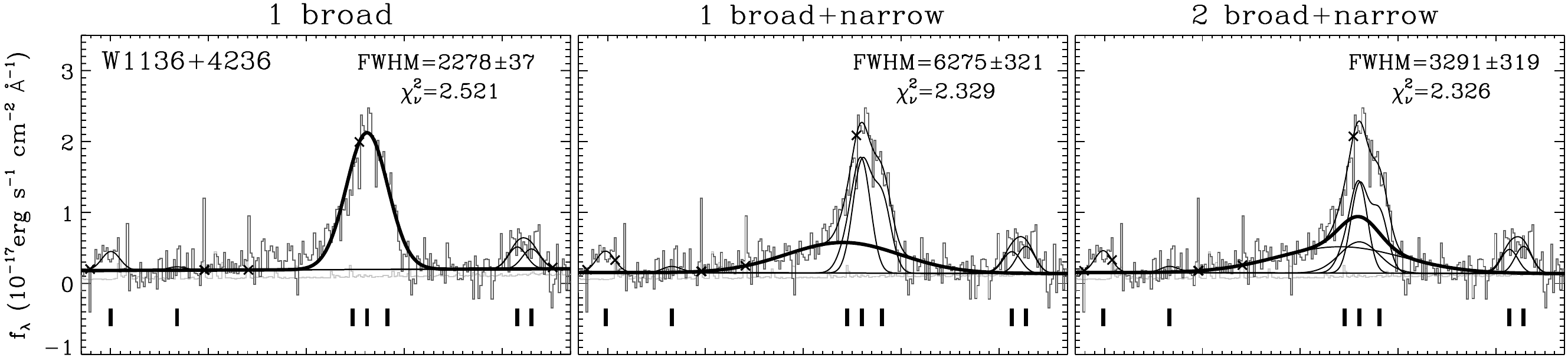}
\plotone{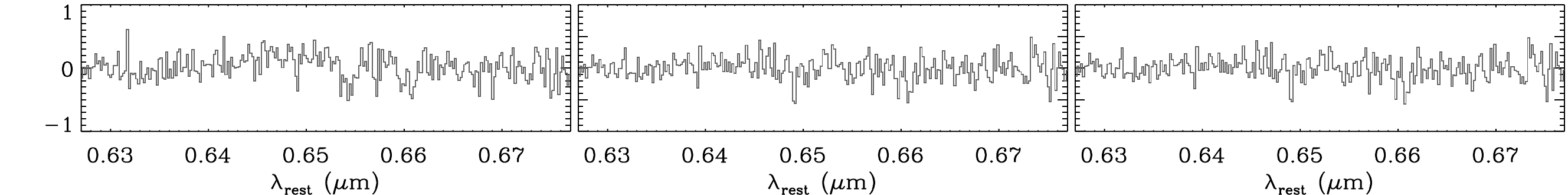}
\plotone{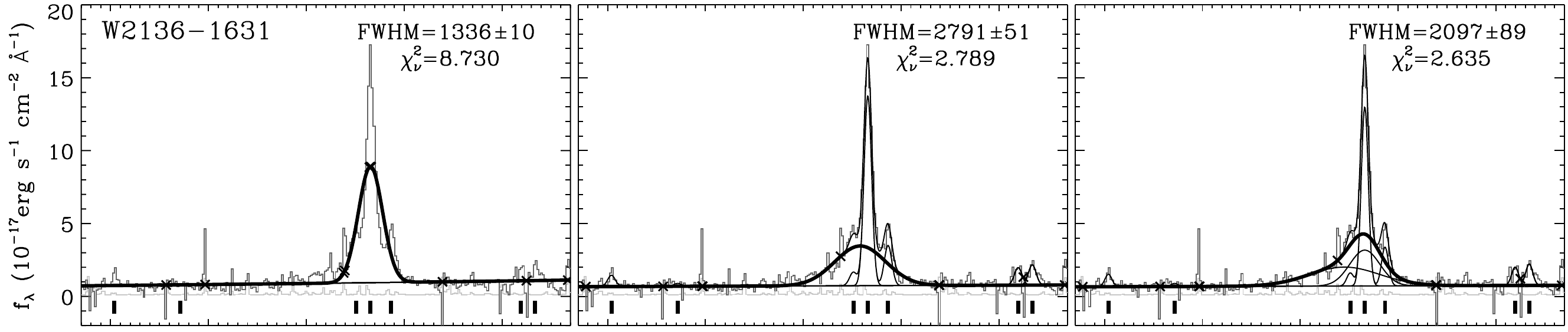}
\plotone{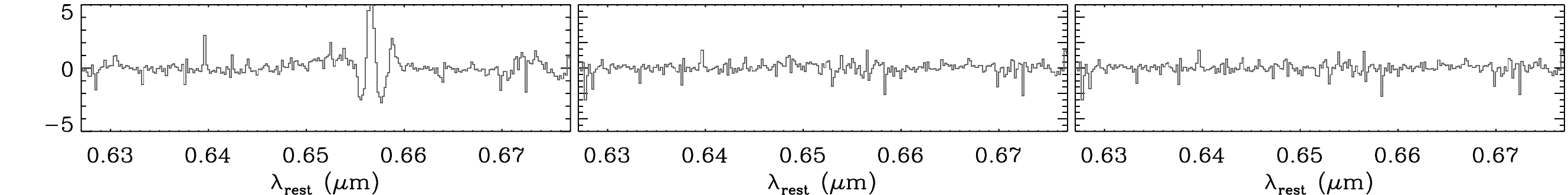}
\plotone{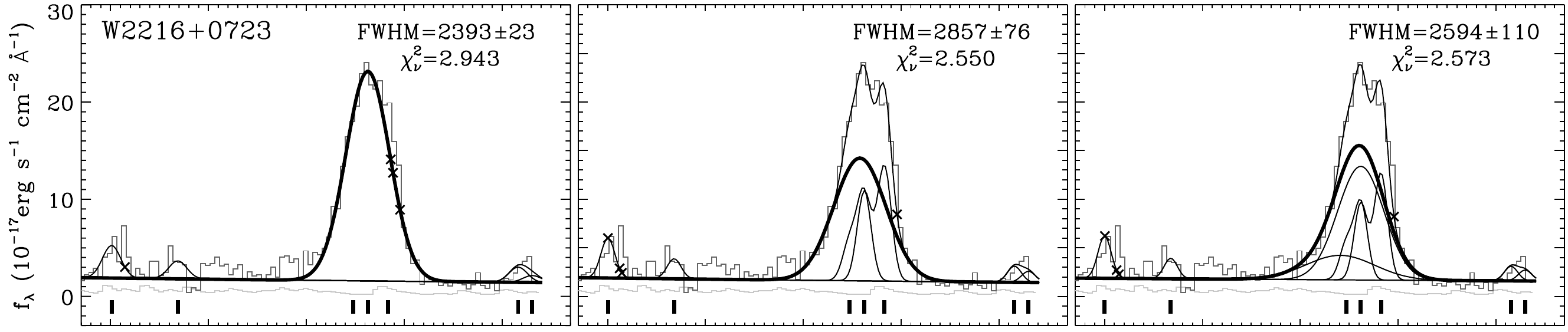}
\plotone{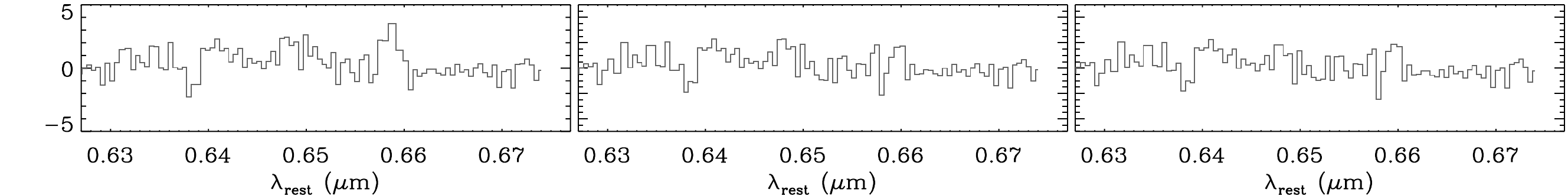}
\plotone{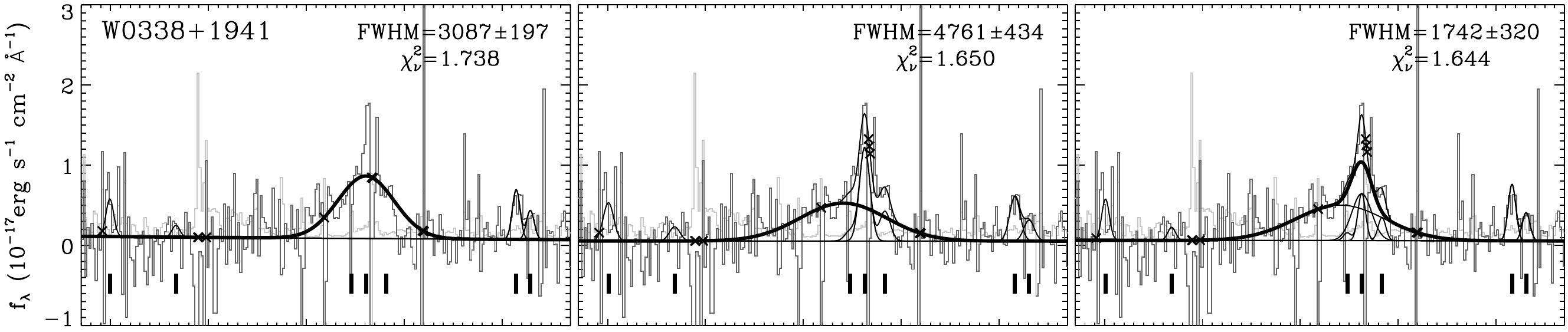}
\plotone{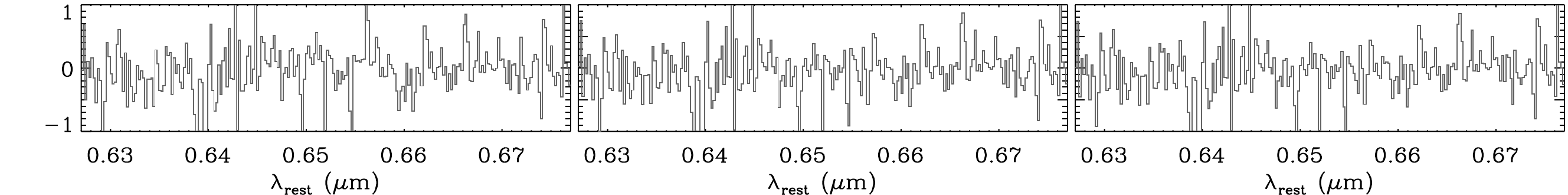}
\plotone{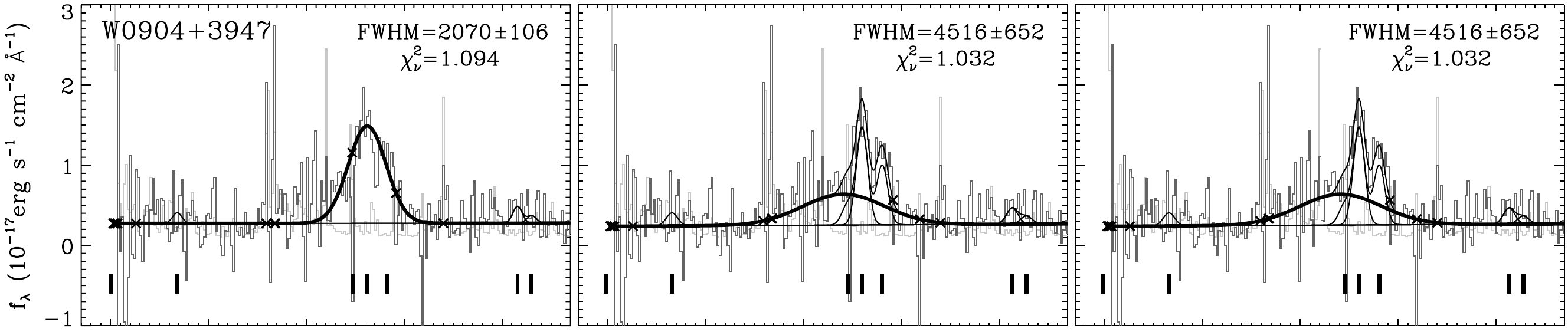}
\plotone{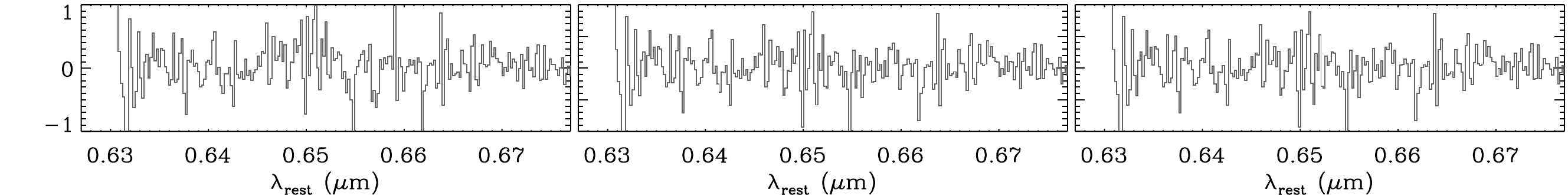}
\caption{$K$-band spectroscopy for three Hot DOGs and $H$-band spectroscopy for two Hot DOGs with model fits. We binned the spectra to their spectral resolution ($\sim$ 3 pixels for MOSFIRE, and $\sim$ 2 pixels for FLAMINGOS-2) for display purposes. Fitted lines are shown in thick black lines, data are plotted in dark grey, and errors are presented in light grey. The ``x" signs mark rejected pixels.  Short vertical lines below the spectra indicate the locations of [OI]$\lambda$6300, 6364, H$\alpha$ $\lambda$6563, [NII]$\lambda$6549, 6585, and [SII]$\lambda$6717, 6732. The residual spectra are shown in the panels below the spectral panels. The three Hot DOGs with good SNR are shown in the top panels.  }
\end{figure*}


\subsection{Can outflows explain the broad H$\alpha$ line?}

Massive outflows have been discovered in some high-redshift, dust-obscured quasars, revealed by the high velocity dispersion of forbidden lines (e.g., Liu et al. 2013a, 2013b; Zakamska et al. 2016),  sometimes accompanied by blue-shifted wings. These features have been explained as a manifestation of strong AGN feedback (e.g., Spoon \& Holt 2009; Mullaney et al. 2013; Zakamska \& Greene 2014; Brusa et al. 2015). In this section, we test if the broad H$\alpha$ lines detected in these Hot DOGs can be explained by outflow combined with emission from NLRs.

We obtained $H$-band spectroscopy for W1136+4236, and detected broad [OIII]$\lambda$4959,5007 lines. We also obtained $J$-band spectroscopy for W2216+0723, with broad [OII]$\lambda$3726,3729 lines detected. The spectra are presented in Figure 2.  Both the [OIII] and [OII]  doublets are thought to be outflow tracers in quasars (e.g., Zakamska \& Greene 2014; Perna et al. 2015).  We fit Gaussian profiles for the [OIII] and [OII] lines. For W1136+4236,  two Gaussians with the same FWHM were fit to the [OIII]$\lambda$4959,5007,  with a FWHM of 2310 km s$^{-1}$. For W2216+0723, we fit a single Gaussian to the [OII]$\lambda$3726,3729  doublet, finding a FWHM of 2607 km s$^{-1}$. Both these forbidden lines are very broad, and present blue-shifted wing features, implying there is outflow in both targets. 

The broad H$\alpha$ we observe in these Hot DOGsÊ can either be caused by virial motion in BLRs right around the massive blackholes, or from more extended NLRs combined with outflows. We made the following tests to see which scenario is preferred by the data. In test 1, we fit a single Gaussian line profile to both the oxygen and H$\alpha$ lines, fixing the linewidth to be the same as [OIII]$\lambda$4959,5007 in W1136+4236, and the same as [OII]$\lambda$3726,3729 in W2216+0723 (Figure 2), assuming the lines are broadened only by outflows. The resulting $\chi^{2}_\nu$s are notably worse than the corresponding models in the 1B+1N and/or 2B+1N models shown in Figure 1. In addition, the resulting line profiles of the [OI] and [SII] lines in this model are too wide to fit the data.Ê In test 2, we used the asymmetric line profile of [OII]$\lambda$3726,3729 to fit H$\alpha$ in W2216+0723 (a similar fit to the [OIII]$\lambda$4959,5007 lines in W1136+4236 did not converge). This led to an even worse $\chi^{2}_\nu$.

In the discussion above, we assume the outflow occurs in the NLRs. This is a plausible scenario to test if outflow from NLRs can cause the observed H$\alpha$ line.  Although we cannot fully exclude the possibility that outflows contribute to the observed broad H$\alpha$, our analysis shows the 1B+1N and 2B+1N models, where the H$\alpha$ lines are broadened by the virial motion from BLRs, provide a better fit to our data than the two test models, where the H$\alpha$ profiles are broadened by NLR outflows.

Even if strong NLR outflows dominate the observed FWHM, preventing us from constraining the BH mass and Eddington ratios in some Hot DOGs, this is still consistent with the picture that Hot DOGs mark a transitional stage in AGN evolution. The detection of very broad [OII] and [OIII] lines is intriguing, as the intense outflows this suggests also imply high accretion rates (e.g. Zakamska et al. 2016). Both are expected for the ``blow-out" phase between the obscured and unobscured quasar stages. While we do not consider BLR outflows in this discussion, we note that strong BLR outflows would likely imply that the virialized BLR component is narrower, which would translate into lower BH masses and, hence, higher, true Eddington ratios. 

\begin{figure*}
\epsscale{0.8}
\plotone{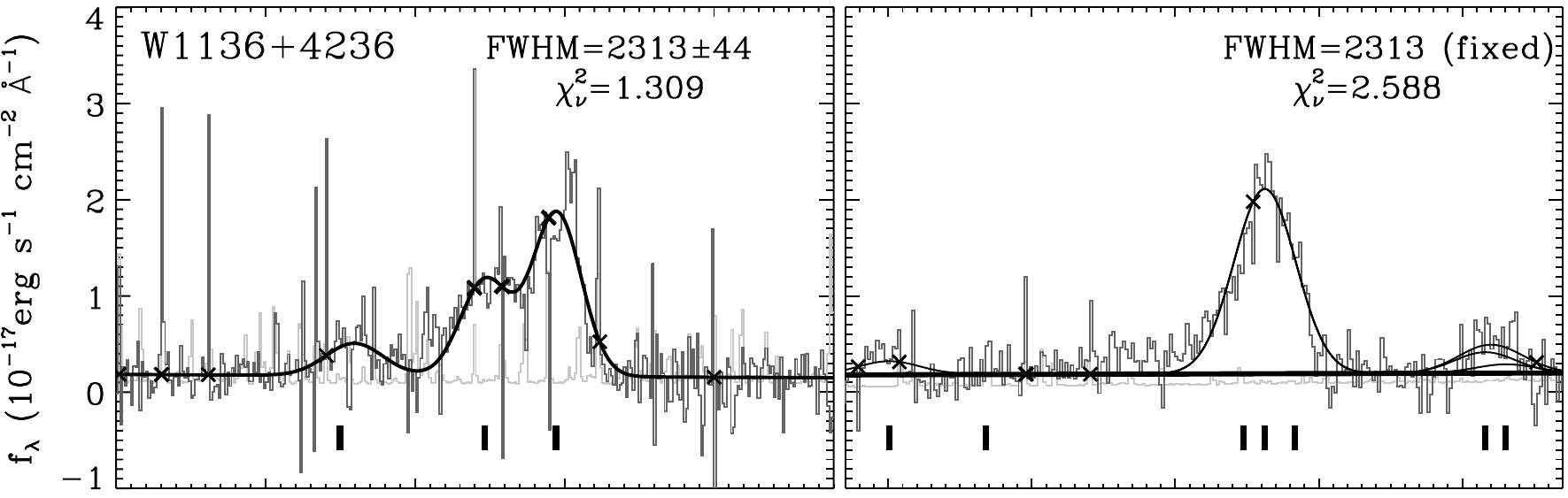}
\plotone{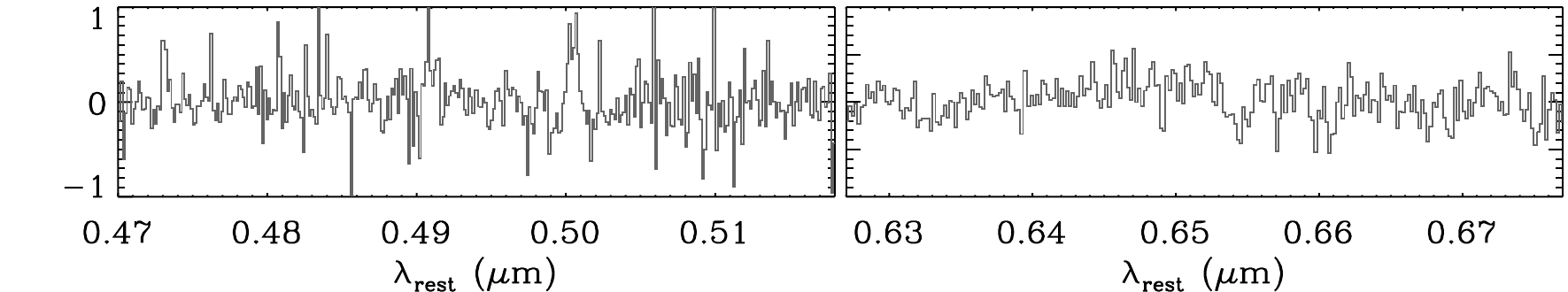}
\plotone{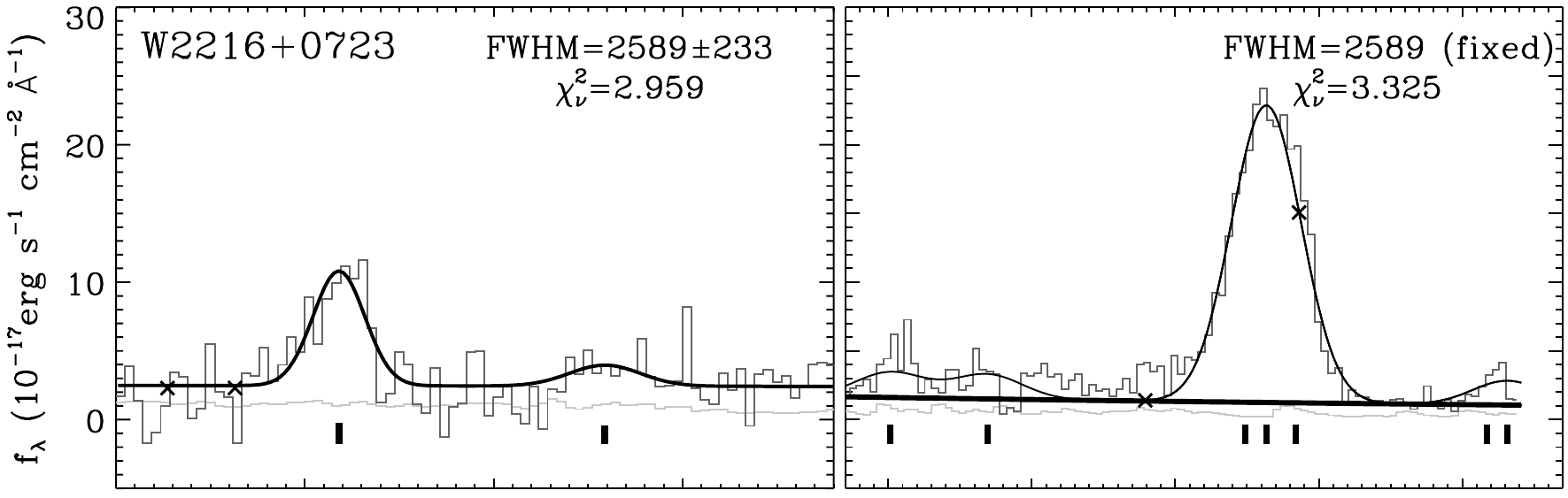}
\plotone{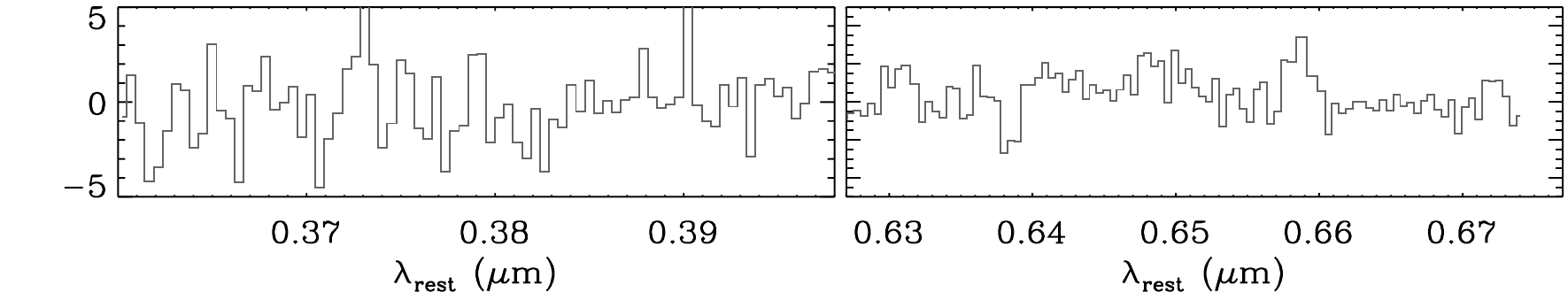}
\caption{Upper panels: $H$- and $K$-band spectra for W1136+4236 with model fits.  We use Gaussians to fit the H$\beta$ and [OIII]$\lambda$4959,5007 lines in the $H$-band spectrum (upper left). In the $K$-band spectrum (upper right), we use Gaussians to fit [OI]$\lambda$6300, H$\alpha$ $\lambda$6563, and [SII]$\lambda$6717, 6732, forcing all linewidths to the same width as [OIII] in the $H$-band. Lower panels: $J$- and $H$-band spectroscopy for W2216+0723 with model fits. We fit one Gaussian for the [OII]$\lambda$3726,3729 lines (lower left), and apply the same line width to fit [OI] and H$\alpha$ in the $H$-band (lower right).  All fitted lines are shown in thick black lines, data are plotted in dark grey, and errors are presented in light grey.  The residual spectra are shown in the panels below the spectral panels. Vertical lines below the spectra show the positions of the H$\beta$ $\lambda$4861 and [OIII]$\lambda$4959,5007 lines in the left panels,  and the [OI]$\lambda$6300, 6364, H$\alpha$ $\lambda$6563, [NII]$\lambda$6549, 6585, and [SII]$\lambda$6717, 6732 lines in the right panels. }

\end{figure*}



\section{Black Hole Masses, Luminosities, and Eddington Ratios} \label{results}

\subsection{Black hole masses}

Black hole mass can be estimated by assuming virial equilibrium in the broad line region: $M_{\rm BH} \propto \it{v}^{2} R_{\rm BLR}/\it{G} $, where $R_{\rm BLR}$ is the radius of the BLR, and {\it{G}} is the gravitational constant. The BLR radius can be measured through the reverberation mapping (RM) technique (e.g., Blandford \& McKee 1982,  Peterson 1993), which estimates the radius of the BLR from the lag between the variability in the AGN continuum and the corresponding variability in the broad permitted lines. This method has been successful in measuring black hole masses in the local universe (e.g., Peterson et al. 2004, Vestergaard \& Peterson 2006). For higher redshift galaxies, BH mass measurements rely on an empirical relation found between the BLR size and the AGN luminosity (the $R-L$ relation) discovered and calibrated by RM studies (i.e. $ R_{BLR} \propto L_{5100}^{0.5}$, where $L_{5100} = \lambda L_{\lambda}$ at $\lambda = 5100 $\AA, Kaspi et al. 2000, Bentz et al. 2006, 2009, 2013), which are called as ``single-epoch" black hole mass measurements. Thus, by measuring rest-frame optical broad line width and adjacent continuum luminosity, one can estimate the single-epoch BH mass.   

 Typically used broad lines to calculate high-redshift AGN BH mass include H$\alpha$, H$\beta$, {\ion{Mg}{2}}, and {\ion{C}{4}}, with H$\alpha$ and H$\beta$ generally regarded as more reliable. The H$\beta$ line has primarily been used to calibrate the $R-L$ relation based on RM measurements, and the H$\alpha$ emission linewidths and luminosities correlate well with those of H$\beta$ over a wide range of total AGN luminosities (e.g., Greene \& Ho 2005; Jun et al. 2015), supporting the use of H$\alpha$ whenever available. Since H$\alpha$ emission is several times stronger than H$\beta$  (Shen \& Liu 2012; Jun et al. 2015), and is less blended with broad {\ion{Fe}{2}} emission than H$\beta$, it is especially preferred for spectroscopy of fainter AGNs. At $z\gtrsim1$, Balmer emission is redshifted out of the optical window,  so rest-frame UV emission ({\ion{Mg}{2}}, {\ion{C}{4}}) has been used to measure BH masses. However, concerns about the scatter of the {\ion{C}{4}}-derived BH masses compared to those derived from Balmer lines (e.g., Assef et al. 2011; Shen \& Liu 2012), as well as strong {\ion{Fe}{2}} blending near {\ion{Mg}{2}}, favor using H$\alpha$ if possible. Moreover, for heavily obscured AGNs like Hot DOGs, UV broad lines are hard to detect. The stronger line and redder wavelength of H$\alpha$ make it a better choice than other lines to probe BLRs in Hot DOGs.
 
Following Assef et al. (2011), we calculate the BH mass based on the FWHM of broad line H$\alpha$: 
$$ M_{\rm BH} = 7.68 \times 10^{6} {\it f} (\frac{FWHM_{H\alpha}}{10^{3} {\rm km\ s^{-1}}})^{2.06} \times (\frac{L_{5100}}{10^{44} {\rm erg\ s^{-1}}})^{0.52} M_{\odot}, $$ 
where $f$ is a scale factor of order unity that depends on the structure, kinematics, and inclination of the BLR (e.g., Collin et al. 2006).  Here we adopt the best-fit fixed $f$ factor of 1.17 following the arguments in Assef et al. (2011). Due to the very high  UV/optical extinction of Hot DOGs, we can't use direct measurements of  $L_{5100}$. Instead, we adopt the rest-frame UV-to-mid-IR SED models for Hot DOGs reported in Assef et al. (2015),  which are constrained by WISE and follow-up optical and near-IR photometry,  fitting the contributions of a starburst, evolved stars, and reddened AGN components, and derive the obscuration-corrected AGN luminosity. The derived extinctions ($A_{V}$) from this model and the resulting BH masses for the five Hot DOGs with broad H$\alpha$ line measurements are presented in Table 4. 

Due to the low SNR for the two H${\beta}$ lines detected in W1136+4236 and W2216+0723, we can't decompose the broad and narrow components for H${\beta}$, complicating the use of the Balmer decrement to estimate extinction. In addition, the possibility that the broad lines are observed due to scattering (Assef et al. 2016) makes the Balmer decrement method potentially problematic. If we assume the same broad to narrow component ratio for H${\alpha}$ and H${\beta}$,  the broad line region extinction for the two sources predicts a much lower AGN luminosity than what we estimated from the observed SED, implying the Balmer decrement method is not a reliable way to estimate the extinction for our sample. 

\subsection{Uncertainties of BH masses}

We estimate the measurement uncertainties in our BH mass calculations by propagating the errors in the FWHM and in the continuum luminosities. The uncertainty in FWHM varies from 8\% to 66\% due to differing data quality. We assume a consistent 50\% uncertainty in calculating $L_{5100}$ from the SED model (Assef et al. 2015). We assume a 20\% uncertainty on the bolometric luminosities (Tsai et al. 2015). The resulting measurement uncertainties are presented in Table 4 and is folded into the error bars in all the Figures.

Some authors (e.g., Jun et al. 2015) also consider measurement errors from the $f$ factor (43\% following Collin et al. 2006) and the $R-L$ relation (7\%,  Bentz et al. 2013). This leads to an additional 0.16 dex uncertainty for BH masses and 0.18 dex uncertainty for Eddington ratios. 
There are also systematic errors for the constants, which is about 0.3-0.4 dex for $f$ factor (Kormendy \& Ho 2013; McConnell et al. 2013) and 0.11 dex for the $R-L$ relation (Peterson et al. 2010).

The uncertainties discussed above do not include the uncertainty introduced by the possible contribution of outflows  to the FWHM of H$\alpha$. As described in Section 3.3, including an outflow component based on the observed [OIII] and [OII] lines for W1136+4236 and W2216+0723, respectively, in modeling their broad H$\alpha$ line profiles leads to a poorer $\chi^{2}_\nu$. To quantitatively estimate the contribution of outflows to FWHM values will require higher signal-to-noise spectra, but is not expected to significantly affect the results presented here.

\subsection{Bolometric luminosities}

Using extensive follow-up photometry (Griffith et al. 2012, Wu et al. 2012, Jones et al. 2014, Assef et al. 2015, Tsai et al. 2015, Tsai et al. in prep.), we constructed complete SEDs for the Hot DOGs and calculated their bolometric luminosities. The method we used is described in detail in Tsai et al. (2015). In brief, we simply integrated over the detected photometric data points from the optical to far-IR bands, using power-law interpolations between measurements and extrapolated to 20\% beyond the shortest and longest wavelengths. This method is more secure than bolometric luminosities simply scaled up from one or a few wavelength measurements using templates, but it is conservative since it may miss flux between data points and beyond the longest/shortest wavelength data. If the best-fit SED templates or spline-smoothed SEDs are considered, the luminosities typically increase by 20\% (Tsai et al. 2015).   

We list the derived bolometric luminosities of the five Hot DOGs in Table 4. The photometry of the Hot DOGs (including the five reported in this paper) observed from {\it Spitzer} (Griffith et al. 2012), WISE and {\it Herschel}, and more details on the method used to calculate their bolometric luminosities (in which we consider  that BH accretion is dominating the luminosity) is reported in Tsai et al. (in prep).

\subsection{Eddington ratios}
The Eddington luminosity is defined as $$ L_{Edd} = 3.28 \times 10^{4} (\frac{ M_{\rm BH}}{M_{\odot}})  L_{\odot} ,$$  and the Eddington ratio is $\eta = L_{AGN}/L_{Edd}$. 
For Hot DOGs, we can attribute most of the bolometric luminosity to the AGN (Eisenhardt et al. 2012; Jones et al. 2014; D\'iaz-Santos et al. 2016, Farrah et al. 2017, Tsai et al. in prep.). Hence we calculated Eddington ratios as $L_{bol}/L_{Edd}$ for the five Hot DOGs (see Table 4). 
We found that $\eta$ from high SNR targets are greater than 0.5 and close to unity. Considering that $L_{bol}$ is calculated conservatively,  we conclude that the derived Eddington ratios are close to or above the Eddington limit for these Hot DOGs.

\section{Discussion } \label{disc}

Based on their SEDs and far-IR to mid-IR luminosity ratios,  Wu et al. (2012) speculated that Hot DOGs may represent a transitional phase following the SMG and DOG phase and preceding the regular quasar phase. Progression along such a sequence is  consistent with galaxy evolution scenarios based on major mergers (e.g., Hopkins et al. 2008) and is likely driven by the growth of the central SMBH.

The primary objective of this study is to measure the BH masses of Hot DOGs, and to test if such an evolutionary sequence is reasonable by comparing their BH masses to those of other populations. The project is also intended to find out if the high luminosities of Hot DOGs result from hosting black holes with masses well above the local BH-host galaxy relation, and/or if their SMBHs are accreting at or even above the Eddington limit. To explore these topics, we compare Hot DOGs to other galaxy populations with measurements of BH masses and Eddington ratios. The discussion is organized as follows: We summarize $z \sim2$ comparison samples and compare them to Hot DOGS in Section 5.1. We discuss possible biases in our results due to our small sample and line fitting approach in Section 5.2. In Section 5.3, we discuss why Hot DOGs are so luminous among quasars, and in Section 5.4 we describe a possible accretion history of $z \sim2$ quasars. In Sction 5.5, we compare Hot DOGs to $z\sim6$ quasars, and propose that Hot DOGs may exist at other redshifts in Section 5.6.

\subsection{Comparison samples at $z \sim2$}

Comparison populations that may relate to Hot DOGs include normal quasars, SMGs, DOGs, and red quasars. Caution should be used when comparing luminosities of obscured and un-obscured populations though, since due to obscuration, some energy output could be missed by scattering to other directions or be reprocessed to unobserved wavelengths. The extinction correction has to be considered when calculating AGN luminosities used to estimate BH masses and Eddington ratios. For Hot DOGs, we argue that  the UV photons have been efficiently converted into mid-IR emission with almost no loss (e.g., Tsai et al. 2015), and we have collected optical to submilimeter SEDs, so that the integrated luminosity should give a good estimate of the bolometric luminosity, and can be compared to total luminosities of other populations.

For normal, unobscured quasars, we use the sample of Shen et al. (2011), who collected redshifts and estimated BH masses for more than 100,000 SDSS quasars from the SDSS DR7 spectroscopic quasar catalog (Abazajian et al. 2009, Schneider et al. 2010). Shen et al. (2011) calculated AGN luminosities by scaling up Type-1 AGN templates using UV continuum measurements, which is reasonable as SDSS quasars generally have low extinction. Their sample covers quasars with redshifts $z \sim 0-5$, and includes measured FWHMs of multiple lines (H$\alpha$, H$\beta$, {\ion{Mg}{2}}, and {\ion{C}{4}}) when they are redshifted into the SDSS spectroscopic range. We are most interested in quasars around $z \sim2$ that can be directly compared to Hot DOGs in this paper. For SDSS quasars, we prefer to use BH masses derived from {\ion{Mg}{2}} rather than from {\ion{C}{4}} over this redshift range, since {\ion{Mg}{2}} is the less complicated line to use as a tool for measuring BH masses (e.g., Shen et al. 2008, 2011, Assef et al. 2011, Jun et al. 2015). 

The BH mass estimate for SMGs is complicated, partly due to the large uncertainty in determining the AGN luminosity in SMGs. Unlike Hot DOGs, the luminosities of SMGs are generally dominated by starbursts rather than AGNs, though some SMGs are actually 850$\mu$m-selected dusty quasars with large extinction. In this paper, we simply adopt average values for general SMGs (whose luminosities are still dominated by starbursts) from Alexander et al. (2008) of log ($M_{\rm BH}$/M$_{\odot}) \sim 8.0$  and $\eta \sim 0.2$.

\begin{figure}
\epsscale{1.2}
\rotatebox{0}{\plotone{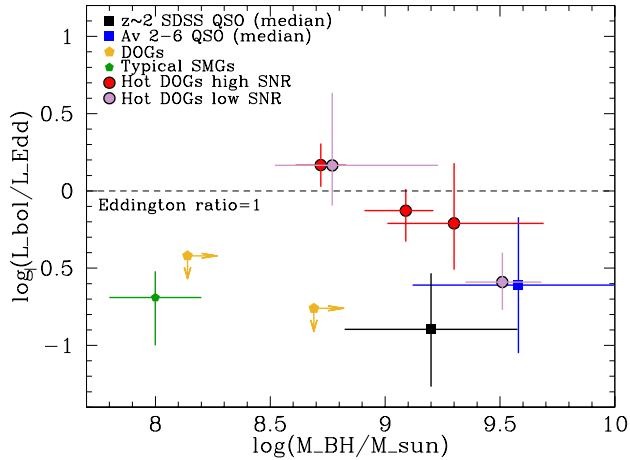}}
\caption{Black hole masses and Eddington ratios for the five Hot DOGs reported in this paper, compared to typical values for $z \sim2$ SMGs (Alexander et al. 2008), DOGs (Melbourne et al. 2011), and quasars (Shen et al. 2011, Banerji et al. 2012, 2015). Note that the Eddington ratios of Hot DOGs are likely 20\% (0.08 dex) higher in general due to our conservative method in calculating bolometric luminosities. The uncertainties of Hot DOGs shown in the figure only consider the measurement errors from the FWHM and the AGN luminosity. An additional error of 0.16 dex on BH mass and 0.18 dex on Eddington ratio should be added if the systematic errors from the $f$ factor and the $R-L$ relations are considered. } 
\end{figure}

Melbourne et al. (2011) report the BH masses of four power-law DOGs based on broad  H$\alpha$ measurements. Two of these (DOG 1 and DOG 4) have reliable bolometric (8-1000 $\mu$m) luminosities from SED integration including  {\it Herschel} data (Melbourne et al. 2012). Since the AGN contribution is believed to dominate the SED of power-law DOGs, we approximate their AGN luminosities using their bolometric luminosities to calculate Eddington ratios, as we did for Hot DOGs.  Melbourne et al. (2011) note that their derived BH masses are lower limits due to the uncertainty in dust correction. In addition, AGNs in DOGs are less dominant than those in Hot DOGs, and the starburst contribution to the luminosity may not be negligible. Therefore, the derived Eddington ratios for these DOGs are upper limits. 
 
There is growing interest in the study of red quasars, which are thought to be a linking population between obscured and unobscured quasars (e.g., Glikman et al. 2012; Ross et al. 2015). Initially, red quasars were selected using optical/near-IR colors (e.g., Richards et al. 2003), and some cross-matching to radio surveys  (e.g., Glikman et al. 2007, 2015).  With the large area, deep mid-IR imaging surveys available from {\it Spitzer} and WISE, additional mid-IR color cuts have been found to be very efficient in selecting red quasars (e.g.,  Banerji  et al.  2013, 2015, Ross et al. 2015, Hamann et al. 2017). These red quasars normally have moderate extinctions ($E(B-V) < 1.5$).
We adopt the luminosities and Eddington ratios from  the red quasar sample of Banerji  et al. (2012, 2015), the largest red quasar sample ($A_{V} \sim 2-6 $) at $z \sim 2$ that has BH masses measured using broad Balmer lines. Like Shen et al. (2011),  Banerji  et al. scaled Type-1 AGN templates using UV continuum measurements to derive bolometric luminosity, but they have corrected the 5100\AA\  luminosity using the obscuration derived from SED fitting.  

In Figure 3, we compare the BH masses and Eddington ratios of Hot DOGs to two DOGs, quasars with moderate obscuration, and Type-1 quasars. A median value is given for each category of quasar. We also mark the likely ranges of BH masses and Eddington ratios for general SMGs reported in Alexander et al. (2008).  The Hot DOGs and comparison samples in Figure 3 have redshifts $z\sim 2$, and BH mass measurements obtained from broad Balmer lines, except for the Type-1 SDSS quasars from Shen et al. (2011).  Figure 3 shows that the Hot DOGs reported here have higher BH masses than general SMGs or DOGs, based on the limited information available, and are comparable to quasars. The Hot DOGs show systematically higher Eddington ratios than other populations at $z\sim2 $, with values close to unity.  Our pilot project supports the idea that Hot DOGs are a transitional, high accreting phase between obscured and unobscured quasars.

\subsection {Biases due to source selection and line fitting}

\subsubsection {Possible selection effects for the pilot sample}

Our pilot project to calculate BH masses and Eddington ratios is based on five Hot DOGs. 
We did not impose any selection criteria apart from redshift, but the small sample size raises the question of how well they represent the BH masses and Eddington ratios for the whole Hot DOG population.

In Figure 4 we compare the distribution of redshifts and luminosities of the five Hot DOGs in this paper to all Hot DOGs with redshifts and {\it Herschel} measurements. Their luminosities are near or somewhat below the median values of Hot DOGs at similar redshifts, and are lower than most Hot DOGs. Hence the Eddington ratios of these five Hot DOGs may be somewhat lower than for typical Hot DOGs with similar BH masses, and it is unlikely there is a selection bias for a high-Eddington ratio subsample because of higher luminosity.

\begin{figure}
\epsscale{1.0}
\plotone{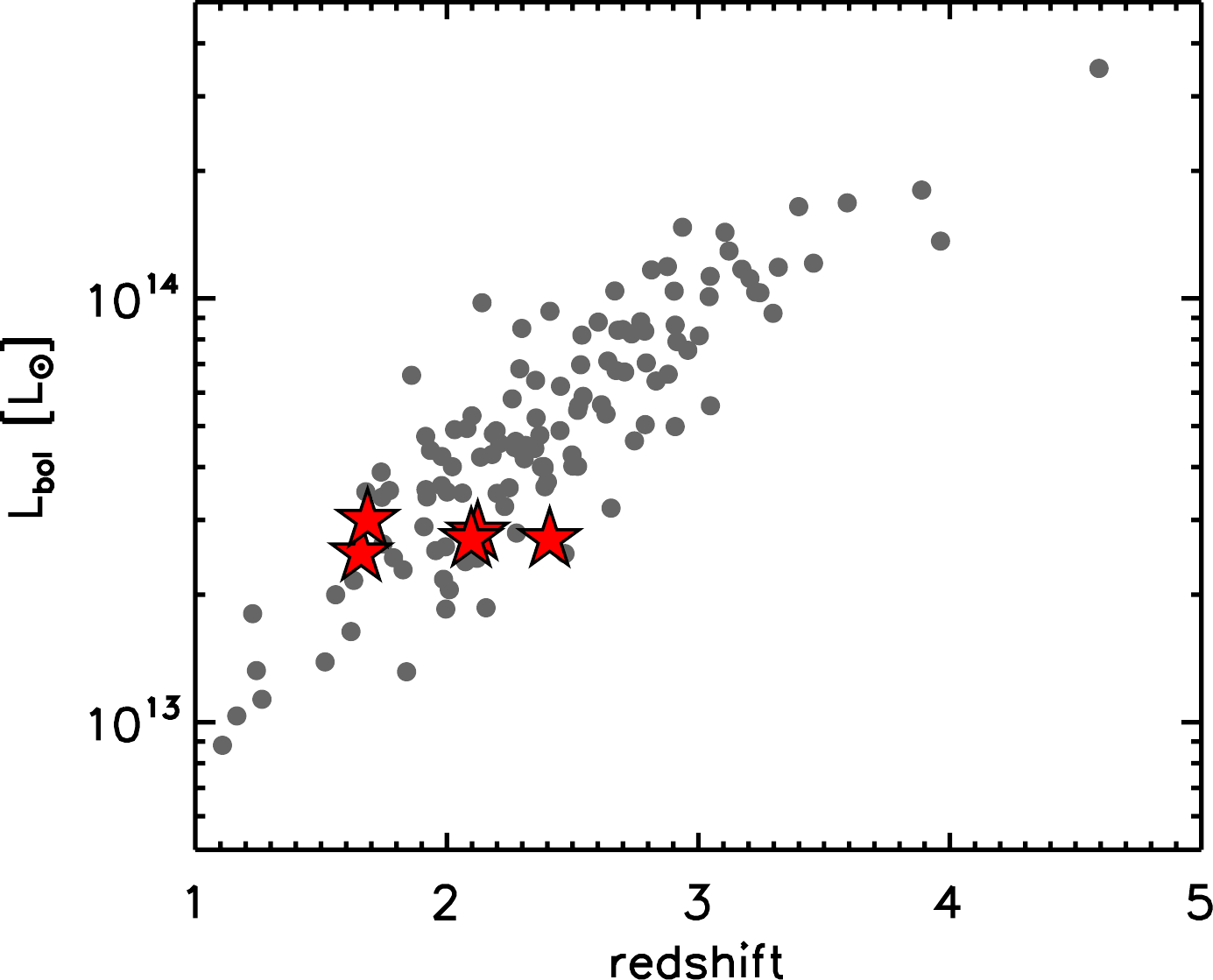}
\caption{The distribution of redshifts and luminosities of the five Hot DOGs reported in this paper (red stars), compared to the sample of Hot DOGs with {\it Herschel} measurements (black dots) presented in Figure 1 of Tsai et al. (2015). }
\end{figure}

Compared with the obscuration values reported in Assef et al. (2015), two of the five targets are close to the lower limit of extinction values, two are close to the median value, and one is close to the upper limit. Overall, there is no obvious bias in extinction.

Although we did not select targets to have broad UV lines, two (W0904+3947 and W1136+4236) out of the five targets do show broad rest-UV spectra features, which may suggest the direct detection of the BLR in the rest-UV. This 40\% ratio is somewhat  higher than the 12\%-16\% fraction of Hot DOGs in general with broad rest-UV lines (Eisenhardt et al. in prep). The relationship between the detection of BLR in rest-optical and rest-UV bands is not clear, but if we assume a higher chance of detecting broad H$\alpha$ for sources with broad rest-UV lines, our pilot sample may be biased against Hot DOGs with narrower or less obvious broad UV-lines. But for a fixed AGN continuum, this means a bias against smaller FWHMs, lower BH masses, and higher Eddington ratios. 

In summary, the possible selection effect of the five Hot DOGs in this paper, including luminosities, obscurations, and existing rest-UV spectra, either have little influence, or suggest even higher Eddington ratios for typical Hot DOGs. These considerations are unlikely to change our conclusion that Hot DOGs represent a high-Eddington ratio population. 

\subsubsection {Influence of different line fitting approaches}

The SNR of our five spectra vary significantly. Our line fitting strategy of using a consistent multi-Gaussian fit takes advantage of the high SNR spectra, and our approach of using spectral information as a prior in Monte-Carlo simulation with an F-test investigation gives a robust and sensitive way to deal with the low SNR spectra and variations between models for some targets. However, some authors prefer to use a simpler approach for lower SNR spectra, namely selecting the simplest Gaussian model with a small enough $\chi^{2}$. How would our results change if we were to take this simpler approach? 

From Table 2 and Figure 1, the simplest Gaussian model with a reasonable $\chi^{2}$ would select the 1B model for all targets except for W2136-1631, for which the 1B+1N model is required. If these models are selected, two targets (W0338+1941, W2136-1631) have larger BH masses and lower Eddington ratios, and the other three targets have lower BH masses and higher Eddington ratios. The range of Eddington ratios hanges from 0.25 -1.46 to 0.45 - 1.31, with the median value changing from 0.74 to 0.88. The overall effect is to move the derived Eddington ratios of Hot DOGs towards unity. Again, this does not change our major conclusion that Hot DOGs are a high-Eddington ratio population.

\subsection {Why are Hot DOGs so luminous: comparing to SDSS quasars }

In the previous section, we showed that Hot DOGs tend to have higher Eddington ratios than other IR luminous (active) galaxy populations atÊ $z \sim2$.Ê Here we argue that an unusually high accretion rate is characteristic of Hot DOGs, and is not a selection effect.Ê In Figure 5, we compare the black hole masses and bolometric luminosities for Hot DOGs to SDSS quasars at all redshifts ($z=0-5$) from Shen et al. (2011), which is the largest and least biased quasar sample with BH masses in the literature.Ê We also include a lower redshift ($z = 1.009$) Hot DOG (WISE J1036+0449) newly reported in Ricci et al. (2017).

\begin{figure*}
\epsscale{1.0}
\plotone{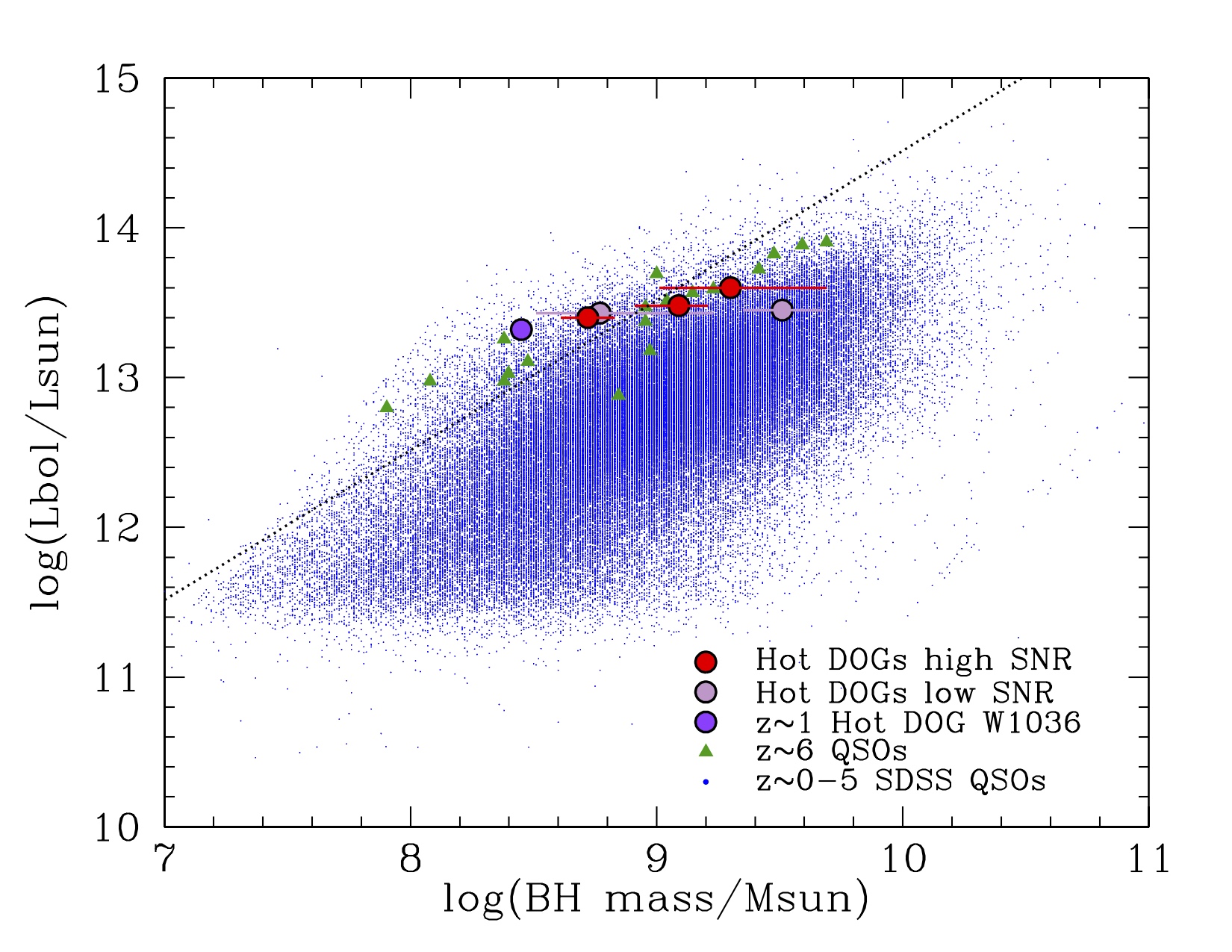}
\caption{Comparison of BH masses and bolometric luminosities of Hot DOGs and SDSS quasars at $z<5$ (Shen et al. 2011) and known $z \sim6$ quasars (Willott et al. 2010, De Rosa et al. 2011). The dotted line traces Eddington ratio =1. Both Hot DOGs and $z \sim 6$ quasars trace the maximum luminosities for quasars at each BH mass. Note that the bolometric luminosity of Hot DOGs in this plot are likely 20\% (0.08 dex) higher in general due to our conservative method in calculating bolometric luminosities. The BH uncertainties of Hot DOGs shown in the figure only consider the measurement errors from the FWHM and the AGN luminosity. An additional error of 0.16 dex should be included If one considers the measurement errors from the $f$ factor and the $R-L$ relations. }
\end{figure*}

Hot DOGs roughly trace the upper luminosity boundary of the SDSS quasar distribution in Figure 5. In other words, they are near the maximum bolometric luminosities seen in quasars with similar BH masses.Ê Such high luminosities must come from very efficient BH accretion. The corresponding Eddington ratios are close to and may even exceed unity (see Figure 3 and Figure 5).  However, the Hot DOGs (especially those with high SNR spectra) are well away from the high mass boundary of the SDSS quasar distribution, arguing against the idea that the high Eddington ratios of Hot DOGs are purely a luminosity selection effect. 

To highlight this further, and reduce possible evolutionary effects,Ê in Figure 6 we Ózoom inÓ to $z =1.65- 1.7$, where two of the three Hot DOGs with high SNR spectra fall. The SDSS quasars at these redshifts display a broad distribution of BH masses extending to both sides of the BH masses for the two Hot DOGs, indicating these Hot DOGs have BH masses typical of normal quasars, but luminosities at the top of the range spanned by SDSS quasars. Figures 4 and 5 imply that Hot DOGs have achieved the highest accretion rates that quasars with similar black hole masses can reach.

\begin{figure}
\epsscale{1.2}
\plotone{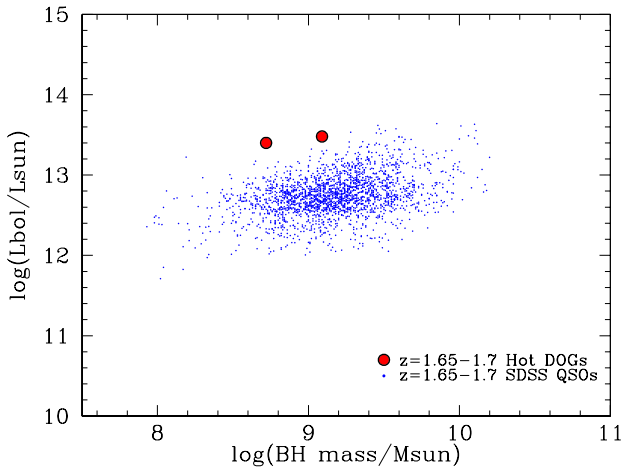}
\caption{Comparison of BH masses and bolometric luminosities of Hot DOGs and SDSS quasars (Shen et al. 2011)   in the redshift range ($1.65 < z < 1.70$).  The bolometric luminosity of Hot DOGs are likely 20\% higher due to our conservative method in calculating bolometric luminosities. }
\end{figure}

In fact, Hot DOGs appear to comprise a significant fraction of the most luminous quasars (Assef et al. 2015) and include the single most luminous galaxy or quasar known (Tsai et al. 2015, D\'iaz-Santos et al. 2016), suggesting that dust obscuration is an important aspect of maximally accreting SMBHs. However, Hot DOGs are not the only populations to reach the highest accretion rates for their BH masses. Some optically selected SDSS quasars have comparably high luminosities and Eddington ratios, as do some very high redshift ($z \sim6$) quasars, as we discuss in Section 5.5.

\subsection {A quasar accretion history at $z \sim2$}

Hot DOGs likely signpost a phase of the highest accretion rate at $z \sim2$, connecting obscured and unobscured quasars. This scenario is consistent with the expectation of a popular galaxy model in which AGN feedback sweeps out the surrounding material that is also the fuel for BH accretion (e.g., Hopkins et al. 2008, Somerville et al. 2008). We should expect to see less obscured quasars associated with larger SMBHs, and with falling accretion rates and fading quasar luminosities. A similar evolutionary connection was also proposed in Assef et al. (2015), based on the luminosity function of Hot DOGs and quasars, that Hot DOGs can be the progenitors of more massive type-1 quasars, in the case that they are experiencing enhanced BH accretion. 

\begin{figure*}
\epsscale{0.8}
\plotone{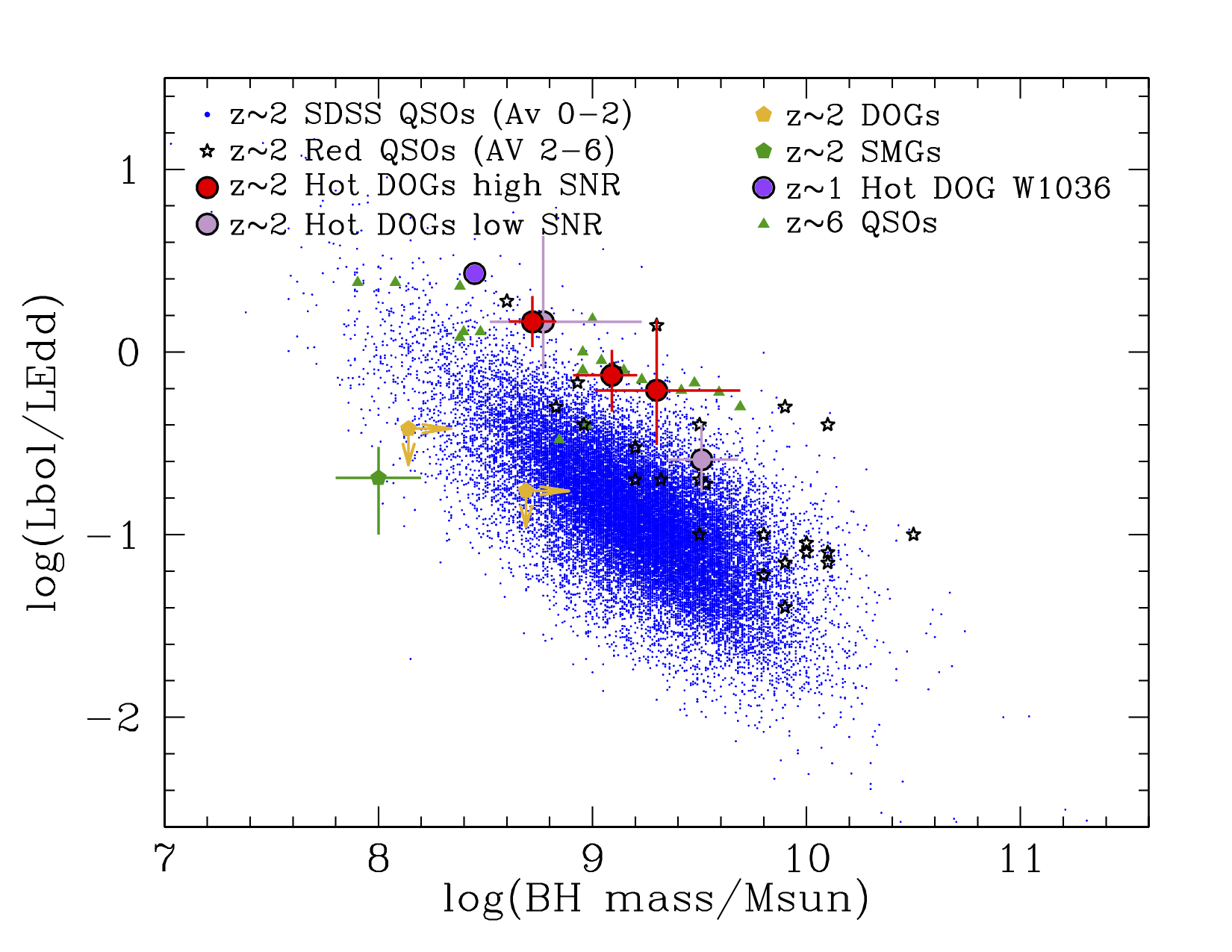}
\caption{Hot DOGs may trace the highest Eddington ratio stage before the red quasar (Banerji et al. 2012, 2015) and SDSS quasar phases (Shen et al. 2011) at  redshift $z\sim 2$. The likely range of BH mass and Eddington ratios for general SMGs (Alexander et al. 2008) and the limits for two DOGs (Melbourne et al. 2011, 2012, Shen et al. 2011) are also marked.  A $z \sim1$ Hot DOG (Ricci et al. 2017) and $z \sim6$ quasars  (Willott et al. 2010, De Rosa et al. 2011) are included for comparison.  Note that the Eddington ratios of Hot DOGs are likely 20\% (0.08 dex) higher due to our conservative method in calculating bolometric luminosities. The uncertainties of Hot DOGs shown in the figure only consider the measurement errors from the FWHM and the AGN luminosity. An additional error of 0.16 dex on BH mass and 0.18 dex on Eddington ratio should be added if the systematic errors from the $f$ factor and the $R-L$ relations are considered. }
\end{figure*}

We can test these predictions using our comparison samples.
Although we do not know how long each evolutionary stage lasts during this evolution, we can take a snapshot of the $z \sim2$ universe to statistically explore the black hole properties for each comparison population introduced in section 5.1. We consider the redshift range $1.5<z<2.5$, which includes the Hot DOGs in this paper.  The peak epoch of quasar density and major merger activities is $z \sim2$, and observational data  at these redshifts is relatively rich.  We focus on BH masses derived from Balmer lines if possible, which are observable from ground-based near-IR spectroscopy at these redshifts. For the SDSS quasars that only have optical spectra, we choose BH masses measured from Mg II instead of C IV, as discussed in Section 5.1. This sets a upper limit of $z=2.25$ for Mg II  to be in the SDSS DR7 spectra. Therefore we have a slightly smaller redshift range of $1.5<z<2.25$ for the SDSS quasar sample, which includes 27761 quasars.

The BH masses and Eddington ratios of Hot DOGs and other comparison samples are plotted in Figure 7. 
Hot DOGs have among the highest Eddington ratios of all quasars with similar black hole masses, much greater than SMGs or DOGs.  Red quasars in Banerji  et al. (2012, 2015) have overlapping but generally lower Eddington ratios than Hot DOGs, but higher Eddington ratios than SDSS quasars, given similar BH masses. The overall average Eddington ratio is  0.37$\pm$0.44 for these $z \sim2$ red quasars, less than Hot DOGs which are close to the Eddington limit.  

Hot DOGs seem to sample not only the highest accretion phase of this evolution, but also are more heavily obscured than other quasars. 
We compare the Eddington ratios and extinction of Hot DOGs, red quasars and regular SDSS quasars in Figure 8. Both Eddington ratios and extinctions roughly follow a decreasing trend along the Hot DOGs, red quasars, and SDSS quasars sequence, suggesting a quasar evolution sequence consistent with the Hopkins et al. (2008) model.

\begin{figure}
\epsscale{1.1}
\plotone{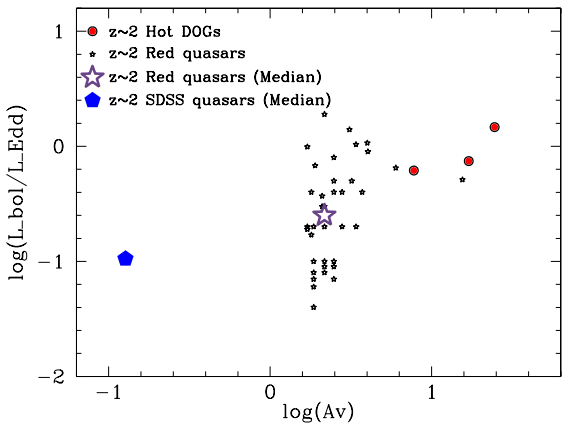}
\caption{Eddington ratio and extinction for $z \sim2$ Hot DOGs, SDSS quasars (Shen et al. 2011), and red quasars (Banerji et al. 2012, 2015). We only include the three Hot DOGs with good SNR spectra. The median values of SDSS quasars and red quasars are marked with blue and purple symbols in the plot.}
\end{figure}

\subsection{Comparison to $z \sim6$ quasars} 

 The consistently high Eddington ratios of $z \sim6$ quasars (e.g., Jiang et al. 2007, Willott et al. 2010, De Rosa et al. 2011) distinguish them from other known SMBH systems, except for the Hot DOG population, as well as a small fraction of high Eddington ratio SDSS quasars at lower redshifts. As shown in Figure 5 and Figure 7, both Hot DOGs and $z \sim6$ quasars have Eddington ratios close to unity, well above most SDSS quasars of similar black hole mass. Hot DOGs have a similar luminosity range ($10^{13} - 10^{14} $L$_{\odot}$), and comparable black hole masses to $z \sim6$ quasars. They are both experiencing high accretion events, in the process of rapidly building up their black holes.

Assef et al. (2015) explores the BH mass vs. host stellar mass ($M_{\rm Sph}$) relation in Figure 7 of their paper, assuming a fixed Eddington ratio of 0.3 that is typical for SDSS quasars. 
They estimated the stellar masses of Hot DOGs by multiplying the rest-frame
luminosity of the host component in the $K$-band by the mass to light ($M/L$) ratio in that band, which depends on many parameters, including star formation history, metallicity, stellar initial mass function,  and the contribution of thermally pulsating AGB stars. They used maximum $M/L$ ratios for these parameters to estimate the upper limits of the stellar masses. Their results suggested Hot DOGs may lie well above the local $M_{\rm BH}-M_{\rm Sph}$ relation. Based on the present work, we can reasonably change the fixed Eddington ratio to unity, and update the $M_{\rm BH}-M_{\rm Sph}$ relation for Hot DOGs, as presented in Figure 9. The Hot DOG stellar masses shown in the plot are again upper limits.\footnote{We also note here that Figure 9 corrects an error in Assef et al. (2015).Ê Assef et al. (2015) stated they used an IMF from Conroy et al. (2013) with a higher $M/L$ ratio in the $K$-band than the $M/L$ for a Chabrier (2003) IMF, but mistakenly used the Chabrier (2013) IMF in their Figure 7. Hence the stellar masses of Hot DOGs shown in Figure 9 in this paper are about a factor of 2 higher than those shown in Figure 7 of Assef et al. (2015).} We mark the three Hot DOGs with high SNR H$\alpha$ spectra as red stars in Figure 9, using the BH masses from this work. After these updates, Hot DOGs are closer to the local $M_{\rm BH}-M_{\rm Sph}$ relation, though these stellar mass estimates are upper limits and their points may move to the left on Figure 9.

Wang et al. (2010) have estimated the BH-host  mass relation for $z \sim6$ quasars, as shown in Figure 9 by green squares. Intriguingly, the region of $z \sim6$ quasars overlaps with Hot DOGs in the $M_{\rm BH}-M_{\rm Sph}$ plot, suggesting they may be hosted by similar kinds of galaxies. The key elements for AGN systems are their black holes masses, host galaxies, and accretion rates. It seems Hot DOGs and $z \sim6$ quasars have similarities in all these elements.

\begin{figure}
\epsscale{1.0}
\rotatebox{0}{\plotone{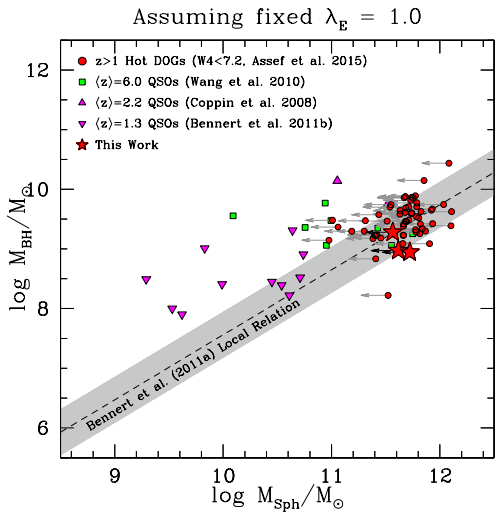}}
\caption{$M_{\rm BH}$ and $M_{\rm Sph}$ values, with Hot DOG data taken from Assef et al. (2015), but assuming a fixed Eddington ratio of 1.0 based on the present work. The high SNR Hot DOG detections from this work are marked with red stars. The bulge masses of host galaxies are constrained by using the best-fit SED template model of Assef et al. (2011, 2015), which are upper limits. A local relation of active galaxies determined by Bennert et al. (2011), as well as the values for quasars at $z \sim6$ (Wang et al. 2010), $z \sim2$ (Coppin et al. 2008) and $z \sim1.3$  (Bennert et al. 2011) are shown for comparison.
}
\end{figure}

The key features of Hot DOGs include high luminosity,  high obscuration, hot dust,  and  high Eddington ratios. Some of these features apply to $z \sim6$\ quasars, too. Significant amounts of hot dust have been revealed in $z \sim6$ quasars (e.g., Wang et al. 2008, Jiang et al. 2010, Leipski et al. 2014). Hot DOGs at $z=2-3$ and quasars at $z \sim6$ may have ultimately been selected for similar reasons: both are experiencing the highest level accretion allowable for their black hole masses, leading to hot dust emission and making both the most luminous objects in their own cosmic epochs.

Quasars at $z \sim6$ are thought to be quite different from lower redshift quasars (e.g., Willott et al. 2010), not only because of their unusually high accretion rates, but also because their surrounding environment could be very different. Quasars at $z \sim6$ represent the earliest generation of SMBHs, when the gas supply was plentiful. Some have argued that  the merger rates were high in the early universe, possibly amenable to spherical accretion (Bondi-Hoyle 1944), and the duty cycle of these SMBHs might have been close to unity (e.g., Willott et al. 2010, Johnson et al. 2012). This is thought to be fundamentally different from lower redshift SDSS quasars, whose SMBHs have passed their peak accretion, and whose host galaxy gas either will not cool or has been cleared out by AGN feedback (e.g., Di Matteo et al. 2008). Thus the duty cycle of lower redshift quasars is much lower. However,  we know little about the immediate surrounding environment of SMBHs at $z \sim6$, except that the host galaxies are dusty.  It is possible that SMBHs in Hot DOGs accrete in a similar way to  $z \sim6$ quasars, at least for a short time, resulting in similar accretion efficiencies for systems with comparable BH and host masses. Whether and how black holes in $z \sim6$ quasars can sustain long-lived Eddington-limited accretion also remains poorly understood. Instead, black holes with multi-episodic, shorter-lived high-accretion rate events are both allowed in simulations reproducing the quasar luminosity function (e.g., Hopkins \& Hernquist 2010), and suggested in some observations (e.g., Jakobsen et al. 2003, Worseck et al. 2007). Models based on disk accretion for $z \sim6$ quasars have also been proposed (e.g., Volonteri, Silk \& Dubus 2015), predicting short-duty-cycle, highly accreting episodes with very high obscuration. 

Compared to $z \sim6$ quasars, Hot DOGs may represent a later generation (or the peak generation of SMBHs) undergoing their maximum accreting episodes. Spectroscopic studies have revealed that emission line properties and metallicity do not evolve much from $z \sim6$ to low redshift quasars (e.g., Dietrich et al. 2003, Fan et al. 2004, De Rosa et al. 2011).  Considering their similarities in BH and host properties, it is possible that the study of Hot DOGs can provide useful constraints and potential insight into understanding the BH accretion and BH-host interaction in $z \sim6$ quasars. 

\subsection{A Hot DOG stage at different cosmic epochs?}

We believe that Hot DOGs are heavily obscured AGNs at a special evolutionary stage, characterized by high luminosity due to high BH accretion rates, and likely with strong AGN feedback. Current work focuses on the studies of $z \sim 2-3$ Hot DOGs. If Hot DOGs trace the transitional, peak accreting phase between obscured and unobscured quasars, should they also exist at other redshift ranges?

Like $z \sim6$ quasars, Hot DOGs are accreting at the maximum level for their BH masses. However, Hot DOGs are heavily dust obscured, while known $z \sim6$ quasars are mostly
selected from their strong rest-frame UV emission (observed at near-IR wavelengths for the most distant systems), and so are more obviously like highly accreting, optically bright $z \sim2-3$ SDSS quasars that may have just passed the accretion peak. Hot DOGs may instead be approaching or at the accretion peak. If there are Hot DOGs at higher redshifts, they would be an obscured version of current known $z \sim6$ quasars. Additional searches using IR photometry available from WISE, {\it Spitzer}, and {\it Herschel} may help to reveal more dusty, higher obscured $z \sim6$ quasars (Blain et al. 2013), with some promising progress made already (e.g., Wu et al. 2015; Carnall et al. 2015; Wang et al. 2016).

There are more obscured AGNs than unobscured ones in the universe at the same bolometric luminosities (e.g., Stern et al. 2012, Assef et al. 2013). Hot DOGs may signpost a key phase linking the well-known unobscured quasar  populations to the lesser known world of obscured quasars. The Hot DOG phase may be followed by a high-luminosity, high-Eddington ratio red/optical quasar stage, present at all redshifts. 

At lower redshifts, Hot DOGs should be rarer due to the decreasing level of quasar activity. Assef et al. (in prep.) have used a revised WISE color criterion to correct the bias of the Hot DOG selection function against $z<2$ objects. One successful example is WISE J1036+0449 at $z=1.009$ (Ricci et al. 2017), whose SED matches well with Hot DOGs, and whose BH mass and Eddington ratio agree with the trends we find in $z \sim2$ Hot DOGs (Figure 5 and Figure 7). For very low redshift quasars, finding a Hot DOG phase object would be even more interesting, since it would provide a rare opportunity to study those maximally accreting stage SMBHs in more detail. For example, the luminous local quasar PDS456 at $z=0.184$ (e.g., Reeves et al. 2000, 2003, Nardini et al. 2015), with a comparable BH mass and luminosity to the $z \sim2$ Hot DOGs in this paper, also accretes at the Eddington limit. This may be a good example of a post-Hot DOG stage quasar in the local universe. More efforts are needed to search for lower redshift Hot DOGs, and here WISE data should prove particularly useful.

\section{Summary} \label{summary}
A population of hyperluminous, dusty galaxies has been discovered by WISE, which we call ``hot, dust-obscured galaxies" or ``Hot DOGs". Their extreme luminosities and hot dust temperatures suggest they either host very massive black holes well above the local BH mass-stellar mass relation, or are accreting at very high rates. We have conducted a pilot survey to measure the BH masses of five Hot DOGs at $z \sim2$, using MOSFIRE at Keck and FLAMINGOS-2 at Gemini. The primary results from this study are summarized below:

1. Broad H$\alpha$ lines were detected in all five targets. Spectral fits imply they are broadened by BLRs around SMBHs. We estimate their BH masses to be $\sim 10^{9} $M$_{\odot}$, and their Eddington ratios are close to unity. 

2. The BH masses are greater than those of typical SMGs and DOGs, and comparable to those of unobscured quasars. This is consistent with the model where Hot DOGs represent a transitional stage between obscured and unobscured quasars. Although not preferred by our spectral fitting, even if strong outflows contribute to the broad line width of H$\alpha$,  the implied strong feedback still supports Hot DOGs' role in the overall evolutionary picture.

3. Hot DOGs have high luminosities compared to quasars with similar black hole masses, which implies they are accreting at the highest possible rates for their SMBH masses,  i.e. they have the highest Eddington ratios observed for quasars and SMBHs. 

4. Our results are consistent with a ``Hot DOG-red quasar-optical quasar" evolutionary sequence.

5. Hot DOGs and $z \sim6$ quasars have comparable BH masses and luminosities, and possibly $M_{\rm BH}-M_{\rm Sph}$ relations. Their SMBHs both accrete at the maximum observed rates, close to the Eddington limit,Ê making them the most luminous persistent objects in their own cosmic epochs.

\acknowledgements
JW is supported by the National Key Program for Science and Technology Research and Development of China (grant 2016YFA0400702) and Project 11673029 supported by NSFC.  HDJ is supported by Basic Science Research Program through the National Research Foundation of Korea (NRF) funded by the Ministry of Education (NRF-2017R1A6A3A04005158). RJA was supported by FONDECYT grant number 1151408. T.D.-S. acknowledges support from ALMA-CONICYT project 31130005 and FONDECYT regular project 1151239. This material is based upon work supported by the National Aeronautics and Space Administration under Proposal No. 13-ADAP13-0092 issued through the Astrophysics Data Analysis Program. This publication makes use of data obtained at the W.M. Keck Observatory, which is operated as a scientific partnership among Caltech, the University of California and NASA. The Keck Observatory was made possible by the generous financial support of the W.M. Keck Foundation. The authors wish to recognize and acknowledge the very significant cultural role and reverence that the summit of Mauna Kea has always had within the indigenous Hawaiian community.  We are most fortunate to have the opportunity to conduct observations from this mountain. This research was partly based on observations obtained at the Gemini Observatory, which is operated by the Association of Universities for Research in Astronomy, Inc., under a cooperative agreement with the NSF on behalf of the Gemini partnership: the National Science Foundation (United States), the National Research Council (Canada), CONICYT (Chile), Ministerio de Ciencia, Tecnolog\'{i}a e Innovaci\'{o}n Productiva (Argentina), and Minist\'{e}rio da Ci\^{e}ncia, Tecnologia e Inova\c{c}\~{a}o (Brazil). This work uses data products from the Wide-field Infrared Survey Explorer, which is a joint project of the University of California, Los Angeles, and the Jet Propulsion Laboratory/California Institute of Technology, funded by the National Aeronautics and Space Administration.  This work uses data obtained from the {\it Spitzer Space Telescope}, which is operated by the Jet Propulsion Laboratory, California Institute of Technology under contract with NASA. This work uses data from {\it Herschel}. {\it Herschel} is an ESA space observatory with science instruments provided by European-led Principal Investigator consortia and with important participation from NASA.

\clearpage{}

\onecolumngrid
\begin{deluxetable}{llllccc}
\tabletypesize{\scriptsize}
\tablenum{1}
\tablewidth{0pt}
\tablecaption{Target list}
\tablehead{
\colhead{Source} &
\colhead{R.A.} &
\colhead{Dec.} &
\colhead{Instrument} &
\colhead{Filter} &
\colhead{UT Date} &
\colhead{Int time (min)}
}

\startdata
W0338$+$1941 & 03:38:51.33 & +19:41:28.6  &    Keck/MOSFIRE  &  $K$ & 2014 Nov 4 & 66    \\
W0904$+$3947 & 09:04:39.84 & +39:47:15.2  &    Keck/MOSFIRE  & $K$  & 2014 Nov 4 & 42 \\
W1136$+$4236 & 11:36:34.29 & +42:36:02.9  &    Keck/MOSFIRE  & $K$, $H$ & 2014 May 6 & 18, 30 \\
W2136$-$1631 & 21:36:55.74 &$-$16:31:37.8 &    Keck/MOSFIRE  & $H$  & 2015 Jun 8 & 48 \\
W2216$+$0723 &  22:16:19.09 & +07:23:53.6   &  Gemini/FLAMINGOS-2 & $J$, $H$  & 2014 Nov 7 & 60
\\         
\enddata

\end{deluxetable}

\begin{deluxetable}{lllllllllll}
\tabletypesize{\scriptsize}
\tablenum{2}
\tablecaption{\label{fitting}Fitting H$\alpha$ with models }
\tablewidth{0pt}
\tablehead{
\multicolumn{1}{l}{} &\multicolumn{3}{c}{1 Broad}&\multicolumn{3}{c}
{1 Broad + 1 Narrow}&\multicolumn{3}{c}{2 Broad + 1 Narrow}\\
\colhead{Source}             &
\colhead{FWHM (km s$^{-1}$)$^{a}$}  &
\colhead{$\chi^{2}_{\nu'}$$^{b}$}  &
\colhead{DOF$^{c}$}    &
\colhead{FWHM (km s$^{-1}$)$^{a}$}  &
\colhead{$\chi^{2}_{\nu'}$$^{b}$}  &
\colhead{DOF$^{c}$}    &
\colhead{FWHM (km s$^{-1}$)$^{a}$}  &
\colhead{$\chi^{2}_{\nu'}$$^{b}$}  &
\colhead{DOF$^{c}$}    
}
\startdata
W0338$+$1941 &     3087(197)  &  1.766   &  271  & 4761(434)  & 1.438   & 268  &  1742(320) & 1.422 & 266 \\
W0904$+$3947 &     2070(106)  &  1.289   & 271   & 4516(652)  & 1.069   & 268   & 4516(652)  & 1.074 & 268 \\
W1136$+$4236 &     2278(37)    & 3.442    &  303   &  6275(321) &  2.618  & 300   & 3291(319) &  2.509 & 298 \\
W2136$-$1631 &     1336 (10)    & 24.716    &  314   &  2791(51)   & 3.845   & 311   & 2097(89)    & 3.199  & 308 \\
W2216$+$0723 &    2393(23)     & 5.529     &  72    &  2857(76)   &  4.067  & 70    & 2594(110)   & 4.178  & 68 \\
\enddata
\tablenotetext{a}{Numbers in parentheses are 1$\sigma$ uncertainties.}
\tablenotetext{b}{ $\chi^{2}_{\nu'}$ are calculated within H$\alpha$ wavelengths (6450-6650~\AA); they are different from $\chi^{2}_{\nu}$ marked in Figure 2 that are calculated over the entire fitting region.}
\tablenotetext{c}{Degrees of freedom in fitting.}

\end{deluxetable}

\begin{deluxetable}{lccc}
\tabletypesize{\scriptsize}
\tablenum{3}
\tablecaption{F-Test result between different spectral fitting models}
\tablehead{
\colhead{Source} &
\colhead{1N vs 1B+1N} &
\colhead{1B vs  1B+1N} &
\colhead{1B+1N vs 2B+1N} 
}

\startdata
W0338$+$1941 &        3.3e-6        &                2.8e-7        &              0.24 \\
W0904$+$3947 &        3.9e-5        &               3.7e-9         &            1.00 \\
W1136$+$4236 &        1.8e-9        &                1.0e-12     &                8.8e-3 \\
W2136$-$1631 &         0                 &               0                  &              1.3e-11 \\
W2216$+$0723 &        3.0e-4        &                9.5e-6       &                0.62  \\       
\enddata

\end{deluxetable}

\begin{deluxetable}{lccccccc}
\tabletypesize{\scriptsize}
\tablenum{4}
\tablecaption{Derived parameters of Hot DOGs}
\tablehead{
\colhead{Source} &
\colhead{$z^{a}$} &
\colhead{$A_{V}$} &
\colhead{FWHM} &
 \colhead{$L_{5100}$}&
\colhead{log(BH mass)$^{b}$} &
 \colhead{L$_{\rm bol}$$^{c}$} &
\colhead{$\eta$}$^{b}$ \\
\colhead{} &
\colhead{} &
\colhead{} &
\colhead{(km s$^{-1}$)} &
 \colhead{(erg s$^{-1}$)} &
\colhead{(log$M_{\odot}$)} &
 \colhead{($L_{\odot}$)} &
\colhead{} 
}

\startdata
W0338$+$1941 &    2.123 & 15.5  & 1742$^{+1144}_{-518}$   &   3.40$\times 10^{46} $ & 8.77$^{+0.46}_{-0.25}$  &   2.8$\times 10^{13}$ & 1.46$^{+2.83}_{-0.66}$ \\
W0904$+$3947      &    2.097 &  8.9   & 4516$^{+812}_{-689}$ &  2.10$\times 10^{46} $ & 9.51$^{+0.17}_{-0.16}$   &   2.7$\times 10^{13}$ & 0.25$^{+0.14}_{-0.08}$\\   
W1136$+$4236             &    2.409 & 7.8    & 3291$^{+1715}_{-1194}$ &  2.85$\times 10^{46} $  & 9.30$^{+0.39}_{-0.29}$    &   2.7$\times 10^{13}$ & 0.62$^{+0.91}_{-0.31}$  \\
W2136$-$1631              &     1.659 & 24.8 & 2097$^{+162}_{-166}$ &  1.30 $\times 10^{46} $  & 8.72$^{+0.11}_{-0.11}$    & 2.5 $\times 10^{13}$ & 1.47$^{+0.55}_{-0.40}$   \\
W2216$+$0723             &   1.685   & 17.1 & 2594$^{+222}_{-497}$ &  2.90$\times 10^{46} $ & 9.09$^{+0.12}_{-0.18}$  & 3.0 $\times 10^{13}$  & 0.74$^{+0.29}_{-0.27}$  
\\         
\enddata
\tablenotetext{a}{Redshifts listed are derived from the spectra presented in this work. The uncertainty of the redshifts is $\Delta z\sim 0.002$.}
\tablenotetext{b}{Uncertainties listed here only consider the measurement uncertainties from the FWHM and the continuum luminosity.} 
\tablenotetext{c}{The uncertainty of the bolometric luminosities is 20\%.}

\end{deluxetable}




\begin{thebibliography}{}

\bibitem[Abazajian et al. (2009)]{Abazajian09}
Abazajian, K. N., Adelman-McCarthy, J. K., Ag\"{u}eros, Marcel A.

\bibitem[Alexander et al. (2008)]{Alexander08}
Alexander, D. M., Brandt, W. N., Smail, I., et al. 2008, AJ, 135, 1968

\bibitem[Antonucci & Miller (1985)]{Antonucci85}
Antonucci, R. R. J., \& Miller, J. S. 1985, ApJ, 297, 621

\bibitem[Antonucci (1993)]{Antonucci93}
Antonucci, R. 1993, Ann. Rev. Ast. \& Astrophys., 31, 473

\bibitem[Assef et al. (2011)]{assef11}
Assef, R. J., Kochanek, C. S., Ashby, M. L. N. et al. 2011, \apj, 728, 56

\bibitem[Assef et al. (2013)]{assef13}
Assef, R. J., Stern, D., Kochanek, C. S., et al. 2013, \apj, 772, 26

\bibitem[Assef et al. (2015)]{Assef15}
Assef, R. J., Eisenhardt, P. R. M., Stern, D., et al. 2015, \aap, 804, 27

\bibitem[Assef et al. (2015)]{Assef16}
Assef, R. J., Walton, D. J.,  Brightman, M., et al. 2016, \aap, 819, 111

\bibitem[Bachetti et al. (2014)]{Bachetti14}
Bachetti, M., Harrison, F. A., Walton, D. J. et al. 2014, Nature, 514, 202

\bibitem[Banerji et al. (2012)]{Banerji12}
Banerji, M., McMahon, R. G., Hewett, P. C., et al. 2012, MNRAS, 427, 2275

\bibitem[Banerji et al. (2013)]{Banerji13}
Banerji, M., McMahon, R. G., Hewett, P. C., et al. 2013, MNRAS, 429, 55

\bibitem[Banerji et al. (2015)]{Banerji15}
Banerji, M.,  Alaghband-Zadeh, S., Hewett, P. C., McMahon, R. G. 2015, MNRAS, 447, 3368

\bibitem[Bennert et al. (2011)]{Bennert11}
Bennert, V. N., Auger, M. W., Treu, T., Woo, J.-H., \& Malkan, M. A. 2011, \aap, 726, 59

\bibitem[Bentz et al. (2006)]{Bentz06}
Bentz, M. C., Peterson, B. M., Pogge, R. W., Vestergaard, M.,  Onken, C. A. 2006, \aap, 644, 133

\bibitem[Bentz et al. (2009)]{Bentz09}
Bentz, M. C., Peterson, B. M., Netzer, H., Pogge, R. W., Vestergaard, M. 2009, \aap, 697, 160

\bibitem[Bentz et al. (2013)]{Bentz13}
Bentz, M. C., Denney, K. D., Grier, C. J. et al. 2013, \aap, 767, 149

\bibitem[Blain et al. (2002)]{blain02}
Blain, A. W., Smail, I., Ivison, R. J., Kneib, J.P., Frayer, D. T. 2002, Physics Reports, 369, Issue 2, 111

\bibitem[Blain et al. (2013)]{blain13}
Blain, A. W., Assef, R.,  Stern, D., et al. 2013, \aap, 778, 113 

\bibitem[Blandford & McKee]{Blandford82}
Blandford, R. D. \& McKee, C. F. 1982, \aap, 255, 419

\bibitem[Bondi & Hoyle (1944)]{Bondi44}
Bondi, H. \& Hoyle, F. 1944, MNRAS, 104, 273

\bibitem[Borys et al. (2005)]{Borys05}
Borys, C., Smail, I., Chapman, S. C., et al. 2005, \aap, 635, 853

\bibitem[Bridge et al. (2013)]{Bridge2013}
Bridge, C., Blain, A., Borys, C. et al. 2013, \aap, 768, 91

\bibitem[Brusa et al. (2015)]{Brusa15}
Brusa, M., Feruglio, C., Cresci, G. et al. 2015, A\&A, 578, 11

\bibitem[Bussmann et al. (2009)]{bussmann09}
Bussmann, R. S., Dey, A., Borys, C., Desai, V., Jannuzi, B. T.,Le Floc\'h, E., Melbourne, J., Sheth, K., Soifer, B. T. 2009, \aap, 705, 184

\bibitem[Bussmann et al. (2015)]{Bussmann15}
Bussmann, R. S., Riechers, D., Fialkov, A., et al.  2015, \aap, 812, 43

\bibitem[Carlstrom et al. (2011)]{Carlstrom11}
Carlstrom, J. E., Ade, P. A. R., Aird, K. A., et al. 2011, PASP, 123, 568

\bibitem[Carnall et al. (2015)]{Carnall15}
Carnall, A. C., Shanks, T., Chehade, B., et al. 2015, MNRAS, 451, 16

\bibitem[Casey et al. (2014)]{Casey14}
Casey, C. M., Narayanan, D., Cooray, A. 2014, Physics Reports, 541, 45

\bibitem[Chapman et al. 2005]{chapman05}
Chapman, S. C., Blain, A., Smail, Ian, Ivison, R. J. 2005, \aap, 622, 772

\bibitem[Chabrier (2003)]{Chabrier03}
Chabrier, G. 2003, PASP, 115, 763

\bibitem[Collin et al. (2006)]{Collin06}
Collin, S., Kawaguchi, T., Peterson, B. M., Vestergaard, M. 2006, A\&A, 456, 75

\bibitem[Conroy et al. (2013)]{Conroy13}
Conroy, C., Dutton, A. A., Graves, G. J., Mendel, J. T., van Dokkum, P. G. 2013, ApJ Letters, 776, 26

\bibitem[Coppin et al. (2008)]{Coppin08}
Coppin, K. E. K., Swinbank, A. M., Neri, R., et al. 2008, MNRAS, 389, 45


\bibitem[De Rosa et al. (2011)]{Derosa11}
De Rosa, G, Decarli, R., Walter, F., 2011, \aap, 739, 56

\bibitem[Dey et al. (2008)]{Dey08}
Dey, A., Soifer, B. T., Desai, V., Brand, K., Le Floc\'{h}, E., Brown, M. J. I., Jannuzi, B. T., Armus, L., Bussmann, S., Brodwin, M., Bian, C., Eisenhardt, P., Higdon, S. J., Weedman, D., Willner, S. P. 2008, \aap, 677, 943

\bibitem[D\'iaz-Santos et al. (2016)]{DiazSantos16}
D\'iaz-Santos, T., Assef, R. J., Blain, A. W. et al. 2016, ApJ Letters, 816, 6

\bibitem[Dietrich et al. (2003)]{Dietrich03}
Dietrich, M., Hamann, F., Appenzeller, I., Vestergaard, M. 2003, \aap, 596, 817

\bibitem[Di Matteo et al. (2008)]{DiMatteo08}
Di Matteo, T., Colberg, J., Springel, V.,  Hernquist, L.,; Sijacki, D. 2008, \aap, 676, 33

\bibitem[Donley et al. (2007)]{Donley07}
Donley, J. L., Rieke, G. H., P\'{e}rez-Gonz\'{a}lez, P. G., Rigby, J. R., Alonso-Herrero, A. 2007, \aap, 660, 167

\bibitem[Dowell et al.(2014)]{Dowell14}
Dowell, C. D., Conley, A., Glenn, J., et al. 2014, \aap, 780, 75

\bibitem[Eikenberry et al. (2012)]{Eikenberry12}
Eikenberry, S., Bandyopadhyay, R., Bennett, J. G. et al. 2012, SPIE, 8446, 0I

\bibitem[Eisenhardt et al. (2012)]{eisenhardt12}
Eisenhardt, P.R.M., Wu, J., Tsai, C. et al.  2012, \aap, 755, 173


\bibitem[Fan et al. (2016a)]{fan16a}
Fan, L., Han, Y., Fang, G., et al. 2016a, \aap, 822, 32

\bibitem[Fan et al. (2016b)]{fan16b}
Fan, L., Han, Y., Nikutta, R. et al. 2016b, \aap, 823, 107

\bibitem[Fan et al. (2004)]{fan04}
Fan, X., Hennawi, J. F., Richards, G. T., et al. 2004, AJ, 128, 515 

\bibitem[Fan et al. (2006)]{fan06}
Fan, X., Carilli, C. L. Keating, B. 2006, Ann. Rev. Ast. \& Astrophys., 44, 415

\bibitem[Farrah et al. (2008)]{08}
Farrah, D., Lonsdale, C. J., Weedman, D. W., Spoon, H. W. W., Rowan-Robinson, M., Polletta, M., Oliver, S., Houck, J. R., Smith, H. E.
2008, \aap, 677, 957

\bibitem[Farrah et al. (2017)]{Farrah17}
Farrah, D., Petty, S et al., Connolly, B. et al. 2017, \apj, 844, 106

\bibitem[Feng & Soria]{feng11}
Feng, H., Soria, R. 2011, New Astronomy Reviews, 55, 166

\bibitem[F\"urst et al. (2016)]{furst16}
F\"urst, F., Walton, D. J., Harrison, F. A. et al. 2016, ApJ, 831, 14

\bibitem[Glikman et al. (2007)]{Glikman07}
Glikman, E., Helfand, D. J., White, R. L., et al. 2007, \aap, 667, 673

\bibitem[Glikman et al. (2012)]{Glikman12}
Glikman, E., Urrutia, T., Lacy, M., et al.  2012, \aap, 757, 51

\bibitem[Glikman et al. (2015)]{Glikman15}
Glikman, E., Simmons, B., Mailly, M., et al.  2015, \aap, 806, 218

\bibitem[Greene & Ho (2005)]{greene05}
Greene, J. E., Ho, L. C.  2005, \aap, 630, 122

\bibitem[Greene et al. (2014)]{Greene14}
Greene, J. E., Alexandroff, R., Strauss, M. A., et al. 2014, \aap, 788, 91 

\bibitem[Griffith et al. (2012)]{Griffith12}
Griffith, R. L., Kirkpatrick, J. D., Eisenhardt, P. R. M., et al. 2012, AJ, 144, 148

\bibitem[Hainline et al. (2014)]{Hainline14}
Hainline, K. N., Hickox, R. C., Carroll, C. M., et al. 2014, \aap, 795, 124

\bibitem[Hamann et al. (2017)]{Hamann17}
Hamann, F., Zakamska, N. L., Ross, N., et al. 2017, MNRAS, 464, 3431 

\bibitem[Hopkins (2006)]{hopkins06}
Hopkins, P F., Hernquist, L., Cox, T. J., Di Matteo, T., Robertson, B., Springel, V. 
2006, \apjs, 163, 1

\bibitem[Hopkins (2008)]{hopkins08}
Hopkins, P. F., Hernquist, L., Cox, T. J., Keres, D. 2008, \apjs, 175, 356

\bibitem[Hopkins & Hernquist (2010)]{hopkinsHernquist10}
Hopkins, P. F, Hernquist, L. 2010, \mnras, 402, 985

\bibitem[Israel et al. (2017a)]{Israel2017a}
Israel, G. L., Papitto, A., Esposito, P., Stella, L. et al.  2017a, MNRAS, 466, 48

\bibitem[Israel et al. (2017b)]{Israel2017b}
Israel, G. L., Belfiore, A., Stella, L. et al. 2017b,  Science, 355, 817

\bibitem[Jakobsen et al. (2003)]{Jakobsen03}
Jakobsen, P., Jansen, R. A., Wagner, S., Reimers, D. 2003, A\&A, 397, 891

\bibitem[Jiang et al. (2007)]{Jiang07}
Jiang, L., Fan, X., Vestergaard, M., et al. 2007, AJ, 134, 1150



\bibitem[Jiang et al. (2010)]{Jiang10}
Jiang, L., Fan, X., Brandt. W. N.,  et al. 2010, Nature, 464, 380

\bibitem[Jones et al. (2014)]{Jones14}
Jones, S. F., Blain, A. W., Stern, D., et al.  2014, MNRAS, 443, 146

\bibitem[Jun & IM]{jun13}
Jun, H. D., Im, M. 2013, \aap, 779, 104

\bibitem[Jun et al. (2015)]{Jun15}
Jun, H. D., Im, M., Lee, H. M., et al. 2015, \aap, 806, 109

\bibitem[Jun et al. (2017)]{Jun17}
Jun, H. D., Im, M.,  Kim, D., Stern , D.  2017, \apj, 838, 41


\bibitem[Kaspi et al. (2000)]{Kaspi00}
Kaspi, S. Smith, P. S., Netzer, H. et al. 2000, \aap, 533, 631

\bibitem[Kleinmann & Low]{Kleinmann70}
Kleinmann, D. E. \& Low, F. J. 1970, \aap, 159, 165

\bibitem[Kormendy \& Ho(2013)]{Kor13} 
Kormendy, J., \& Ho, L.~C.\  2013, Ann. Rev. Ast. \& Astrophys., 51, 511

\bibitem[Leipski et al. (2014)]{Leipski14}
Leipski, C., Meisenheimer, K., Walter, F. et al. 2014, \aap, 785, 154

\bibitem[Liu et al. (2013)]{Liu13}
Liu, G., Zakamska, N. L., Greene, J. E., Nesvadba, N. P. H., Liu, X. 2013, MNRAS, 436, 2576

\bibitem[Liu et al. (2014)]{Liu14}
Liu, G., Zakamska, N. L., Greene, J. E. 2014, MNRAS, 442, 1303

\bibitem[Lonsdale et al. (2009)]{Lonsdale09}
Lonsdale, C. J. et al. 2009, \apj, 692, 442

\bibitem[Makishima et al (2000)]{Makishima00}
Makishima, K., Kubota, A., Mizuno, T. et al. 2000, \aap, 535, 632 

bibitem[Magnelli et al. (2012)]{Magnelli2012}
Magnelli, B., Lutz, D. Santini, P. et al. 2012, A\&A, 539, 155

bibitem[Markwardt (2009)]{Markwardt09}
Markwardt, C. B.  2009, ASPC, 411, 251

\bibitem[McConnell \& Ma(2013)]{McC13} 
McConnell, N.~J., \& Ma, C.-P.\  2013, ApJ, 764, 184

\bibitem[Melbourne et al. (2011)]{Melbourne2011}
Melbourne, J., Peng, C. Y., Soifer, B. T. et al. 2011, AJ, 141, 141

\bibitem[Melbourne et al. (2012)]{Melbourne2012}
Melbourne, J., Soifer, B., Desai, V. et al. 2012, AJ, 143, 125

\bibitem[Miller & Colbert (2004)]{miller04}
Miller, M. C. \& Colbert, E. J. M.,  2004, International Journal of Modern Physics, 13, 1

\bibitem[Mullaney et al. (2013)]{Mullaney13}
Mullaney, J. R., Alexander, D. M., Fine, S., Goulding, A. D., Harrison, C. M., et al.  2013, MNRAS, 433, 622

\bibitem[Narayanan et al. (2010)]{narayanan10}
Narayanan, D., Dey, A., Hayward, C. C., Cox, T.J., Bussmann, R. S., Brodwin, M., Jonsson, P., Hopkins, P. F., et al.  2010, MNRAS, 407, 1701

\bibitem[Nardini et al. (2015)]{Nardini15}
Nardini, E., Reeves, J. N., Gofford, J., et al. 2015, Science, 347, 860

\bibitem[Negrello et al. (2010)]{Negrello10}
Negrello, M., Hopwood, R., De Zotti, G., et al. 2010, Science, 330, 800 

\bibitem[Neugebauer et al. (1984)]{Neugebauer84}
Neugebauer, G., Habing, H. J., van Duinen, R., et al.  1984, ApJ Letters, 278, 1


\bibitem[Ohsuga et al. (2002)]{Ohsuga02}
Ohsuga, K., Mineshige, S., Mori, M., Umemura, M. 2002, \aap, 574, 315

\bibitem[Ohsuga et al. (2005)]{Ohsuga05}
Ohsuga, K., Mori, M., Nakamoto, T., Mineshige, S. 2005, \aap, 628, 368

\bibitem[Peterson (1993)]{Peterson93}
Peterson, B. M. 1993, PASP, 105, 24

\bibitem[Peterson et al. (2004)]{Peterson04}
Peterson, B. M., Ferrarese, L., Gilbert, K. M., et al. \aap, 613, 682

\bibitem[Peterson (2010)]{Peterson10}
Peterson, B. M. 2010, in IAU Symp. 267, Co-Evolution of Central Black
Holes and Galaxies, ed. B. Peterson, R. Somerville, \& T. Storchi-Bergmann
(Cambridge: Cambridge Univ. Press), 151

\bibitem[Piconcelli et al. (2015)]{Piconcelli15}
Piconcelli, E., Vignali, C., Bianchi, S., et al. 2015, A\&A, 574, 9

\bibitem[Pilbratt et al. (2010)]{Pilbratt10}
Pilbratt, G. L., Riedinger, J. R., Passvogel, T., et al. 2010, A\&A, 518, 1

\bibitem[Reeves et al. (2000)]{Reeves00}
Reeves, J. N., O'Brien, P. T., Vaughan, S., et al.  2000, MNRAS, 312, 17

\bibitem[Reeves et al. (2003)]{Reeves03}
Reeves, J. N., O'Brien, P. T., Ward, M. J. 2003, \aap, 593, 65

\bibitem[Ricci et al. (2017)]{Ricci17}
Ricci, C., Assef, R. J., Stern, D., et al. 2017, \aap, 835, 105

\bibitem[Richards et al. (2003)]{Richards03}
Richards, G. T., Hall, P. B., Vanden Berk, D. E. et al. 2003, AJ, 126, 1131

\bibitem[Riechers et al. (2013)]{Riechers13}
Riechers, Dominik A.; Bradford, C. M.; Clements, D. L. et al. 2013, Nature, 496, 329

\bibitem[Rieke & Low (1972)]{Rieke72}
Rieke, G. H. \& Low, F. J. 1972, 1972, ApJ Letters, 176, 95

\bibitem[Rigby et al. (2004)]{Rigby04}
Rigby, J. R., Rieke, G. H., Maiolino, R. et al. 2004, ApJS, 154, 160.


\bibitem[Ross et al. (2015)]{Ross15}
Ross, N. P., Hamann, F., Zakamska, N. L., et al. 2015, MNRAS, 453, 3932


\bibitem[Schneider et al. (2010)]{Schneider10}
Schneider, D. P., Richards, G. T., Hall, P. B., et al. 2010, AJ, 139, 2360


\bibitem[Shen et al. (2008)]{shen08}
Shen, Y., Greene, J. E., Strauss, M. A., Richards, G. T., Schneider, D. P. \aap, 680, 169

\bibitem[Shen et al. (2011)]{shen11}
Shen, Y., Richards, G. T., Strauss, M. A. et al. 2011, \aap, 194, 45

\bibitem[Shen & Liu (2012)]{shen12}
Shen, Y. \& Liu, X. 2012, \aap, 753, 125

\bibitem[Shen & Ho(2014)]{shen14}
Shen, Y., Ho, L. C. 2014, Nature, 513, 210

\bibitem[Somerville et al. (2008)]{Somerville08}
Somerville, R. S., Hopkins, P. F., Cox, T. J., Robertson, B. E. Hernquist, L. 2008, MNRAS, 391, 481

\bibitem[Spoon & Holt (2009)]{spoon09}
Spoon, H. W. W. \& Holt, J. 2009, ApJ Letters, 702, 42

\bibitem[Stern et al. (2012)]{Stern12}
Stern, D., Assef, R. J., Benford, D. J., et al. 2012, \aap, 753, 30

\bibitem[Stern et al. (2014)]{Stern14}
Stern, D., Lansbury, G. B., Assef, R. J., et al. 2014, \aap, 794, 102

\bibitem[Swetz et al. (2011)]{Swetz11}
Swetz, D. S., Ade, P. A. R., Amiri, M., et al. 2011, ApJS, 194, 41 

\bibitem[Trump et al. (2011)]{trump11}
Trump, J. R., Impey, C.D., Kelly, B. C. 2011, \aap, 733, 60

\bibitem[Tsai et al. (2015)]{Tsai15}
Tsai, C.,  Eisenhardt, P. R. M., Wu, J., et al. 2015, \aap, 805, 90

\bibitem[Vacca et al. (2003)]{Vacca03}
Vacca, W. D., Cushing, M. C., Rayner, J. T. 2003, PASP, 115, 389

\bibitem[Vestergaard & Peterson (2006)]{Vestergaard06}
Vestergaard, M. \& Peterson, B. M. 2006, \aap, 641, 689

\bibitem[Vieira et al. (2013)]{Vieira13}
Vieira, J. D., M. D. P.; Chapman, S. C., et al. 2013, Nature, 495, 344

\bibitem[Volonteri et al. (2015)]{Volonteri15}
Volonteri, M., Capelo, P. R., Netzer, H., et al. 2015, MNRAS, 452, 6

\bibitem[Wang et al. (2015)]{Wang15}
Wang, F., Wu, X., Fan, X., et al. 2015, ApJ Letters, 807, 9

\bibitem[Wang et al. (2008)]{wang08}
Wang, R., Carilli, C. L., Wagg, J., et al. 2008, \aap, 687, 848

\bibitem[Wang et al. (2010)]{wang10}
Wang, R., Carilli, C. L., Neri, R. et al., 2010, \aap, 714, 699




\bibitem[Willott et al. (2010)]{willott10}
Willott, C. J., Albert, L., Arzoumanian, D. et al.,  2010,  AJ, 140, 546

\bibitem[Worseck et al. 92007)]{Worseck07}
Worseck, G., Fechner, C., Wisotzki, L., Dall'Aglio, A, 2007, A\&A, 473, 805

\bibitem[Wright et al. (2010)]{Wright10}
Wright, E. L., Eisenhardt, P. R. M., Mainzer, A. K., et al. 2010, AJ, 140, 1868

\bibitem[Wu et al. (2012)]{wu12}
Wu, J., Tsai, C., Sayers, J., et al., 2012, ApJ, 756, 96

\bibitem[Wu et al. (2014)]{wu14}
Wu, J., Bussmann, R. S., Tsai, C., et al. 2014, \aap, 793, 8

\bibitem[Wu et al. (2015)]{wu15}
Wu, X.,  Wang, F., Fan, X., et al. 2015, Nature, 518, 512

\bibitem[Yan et al. (2007)]{yan07}
Yan, L., Sajina, A., Fadda, D., Choi, P., Armus, L., Helou, G., Teplitz, H, Frayer, D., Surace, J. 2007, \apj, 658, 778

\bibitem[Zakamska & Greene]{Zakamska14}
Zakamska, N. L., \& Greene, J. E. 2014, MNRAS, 442, 784

\bibitem[Zakamska et al. (2016)]{Zakamska16}
Zakamska, N. L., Hamann, F., P\^{a}ris, I., et al. 2016, MNRAS, 459, 3144

\end{thebibliography}

\end{document}